
\input amstex
\documentstyle{amsppt}
\NoRunningHeads
\NoBlackBoxes
\topmatter
\title INFINITE DIMENSIONAL GEOMETRY AND\linebreak
QUANTUM FIELD THEORY OF STRINGS. III\linebreak
INFINITE DIMENSIONAL W-GEOMETRY OF\linebreak
A SECOND QUANTIZED FREE STRING.
\endtitle
\author {\tt D.V.JURIEV}\footnote{\rm On leave from Mathematical Division of
Research Institute for System Studies, Moscow, Russia (e mail:
juriev\@systud.msk.su)
\newline}
{\it \centerline{Laboratoire de Physique Th\'eorique}
\centerline{de l'\'Ecole Normale Sup\'erieure,}\newline
24 rue Lhomond, 75231 Paris Cedex 05, France
\footnote{\rm Unit\'e Propre du Centre National de la Recherche
Scientifique associ\'ee \`a l'\'Ecole Normale
Sup\graveaccent erieure et \`a l'Universit\'e de Paris-Sud\newline}}
\linebreak
\centerline{\rm E mail: juriev\@physique.ens.fr}
\ \newline
\endauthor
\abstract The present paper is devoted to various objects of the
infinite dimensional $W$-geometry of a second quantized free string.
Our purpose is to include the $W$-symmetries into the general infinite
dimensional geometrical picture related to the quantum field theory of strings,
which was described in [1]. It is done by a change of the Lie algebra of all
infinitesimal reparametrizations of a string world-sheet on the Lie
quasi(pseudo)algebra of classical $W$-transformations (Gervais-Matsuo
quasi(pseudo)algebra) as well as of the Virasoro algebra on the central
extended enlarged Gervais-Matsuo quasi(pseudo)algebra. A way to obtain
$W$-algebras from the classical $W$-transformations (i.e. Gervais-Matsuo Lie
quasi(pseudo)algebra) is proposed. The relations of Gervais-Matsuo
differential $W$-geometry to the Batalin-Weinstein-Karasev-Maslov approach to
the geometry of nonlinear Poisson brackets as well as to L.V.Sabinin program
of "nonlinear geometric algebra" are mentioned.
\endabstract
\endtopmatter
\document
\define\ssl{\operatorname{sl}}
\define\FG{\operatorname{FG}}
\define\id{\operatorname{id}}
\define\SU{\operatorname{SU}}
\define\Rot{\operatorname{Rot}}
\define\BP{\operatorname{BP}}
\define\SI{\operatorname{SI}}
\define\enl{\operatorname{enl}}
\define\gl{\operatorname{gl}}
\define\sgrad{\operatorname{sgrad}}
\define\Ddiv{\operatorname{div}}
\define\Ddet{\operatorname{det}}
\define\Aut{\operatorname{Aut}}
\define\SltZ{\operatorname{SL}(2,\Bbb Z)}
\define\tor{\Bbb T^2}
\define\plane{\Bbb R^2}
\define\circle{\Bbb S^1}
\define\Symp{\operatorname{Symp}}
\define\Ham{\operatorname{Ham}}
\define\CVect{\operatorname{\Bbb CVect}}
\define\Cvir{\operatorname{\Bbb Cvir}}
\define\DOP{\operatorname{DOP}}
\define\PDOP{\operatorname{PDOP}}
\define\LDOP{\DOP_{[\cdot,\cdot]}}
\define\LPDOP{\PDOP_{[\cdot.\cdot]}}
\define\Tr{\operatorname{Tr}}
\define\Res{\operatorname{Res}}
\define\Moy{\operatorname{Moy}}
\define\cLDOP{\widehat{\DOP}_{[\cdot,\cdot]}}

\define\GD{\operatorname{GD}}
\define\Ker{\operatorname{Ker}}
\define\Iim{\operatorname{Im}}
\define\Rad{\operatorname{Rad}}
\define\Md{\operatorname{Md}}
\define\LMd{\Md_{[\cdot,\cdot]}}
\define\RW{\operatorname{\Cal R\Cal W}}
\define\cRW{\widehat{\RW}}
\define\sltC{\operatorname{sl}(2,\Bbb C)}
\define\slnC{\operatorname{sl}(n,\Bbb C)}
\define\oddim{\operatorname{oddim}}
\define\ad{\operatorname{ad}}
\define\Diff{\operatorname{Diff}}
\define\Vir{\operatorname{Vir}}
\define\Map{\operatorname{Map}}
\define\sq{\cdot}
\define\Gr{\operatorname{Gr}}
\define\Fl{\operatorname{Fl}}
\define\CP{\operatorname{\Bbb C\Bbb P}}
\define\reg{\operatorname{reg}}
\define\EXP{\operatorname{EXP}}
\define\CGM{\operatorname{\Cal G\Cal M}}
\define\GM{\operatorname{GM}}
\define\NG{\operatorname{NG}}
\define\BRST{\operatorname{BRST}}
\define\Vect{\operatorname{Vect}}
\document
\head Introduction \endhead
This is the third part of the text devoted to an infinite dimensional
geometry of a quantum field theory of strings (for previous ones see [1,2]).
It is devoted to an infinite dimensional geometry, which appears in the theory
of second quantized free strings, as well as the first one.
An origin of the writing of this article is hidden in some very crucial detail
of the theory, which is known under the title of $W$-symmetries.
The algebraic aspects of $W$-symmetries, cumbersome and mysterious a little
at the beginning [3], were recently clarified by B.A.Khesin, V.Yu.Ovsienko,
A.O.Radul et al [4-10].
Moreover, a simple geometric interpretation of them was given by J.-L.Gervais
and Y.Matsuo [11].
It seems that the cruciality of $W$-symmetries to the theory of second
quantized free strings is in the fact that $W$-"reparametrizations" of strings,
which are considered, may be not obligatory global in the loop space
(i.e.unique for all strings independently on their position in a target space)
but in some sense local, i.e. depends in some natural way on an embedding of
the string into the target space.
One may say that such transformations are related not only to the intrinsic
but also extrinsic geometry of a string.
A class of the most interesting transformations of this kind, which are related
to a certain complex analog of a classical Frenet theory of curves invariants
[12], was described by J.-L.Gervais and Y.Matsuo as classical
$W$-transformations.
Such transformations of both internal and external degrees of friedom of
a string maybe considered as hidden ones with respect to standard conformal
symmetries.
So a natural question arizes: how the process of a geometric quantization of a
string changes under an account of these hidden symmetries.
{}From the mathematical point of view it may be considered as a problem of
an "induction" of the quantization process directed by a considered symmetry
algebra to one directed by its extension.
So the solution of this problem (reproduced from a typical pattern for
this partial infinite dimensional case of the quantum field theory of free
strings, which we are considered) should produce a lot of new geometric
material, completing a general picture, which was drawn in [1], by
new intriguing details and nuances.

This opinion is also confirmed by the fact that $W_\infty$-symmetries are very
deeply
related to the group of area-preserving diffeomorphisms of a torus.
This group, the corresponding Lie algebra, its deformations and central
extensions, being the main objects of a certain development of the string
theory, the membrane theory, produce the next scope of infinite dimensional
objects after one of the Virasoro algebra.
So we may consider a geometric material, which will be discussed in the paper,
as a possible introduction to the transition from the careful consideration of
the infinite dimensional geometry of objects from the "Virasoro family", which
was begun by A.A.Kirillov and the author in [13] and summarized partially
in [14,1,2] (see also references wherein) to a systematic treatment of
geometric material, concerning objects related to the group of
symplectomorphisms of a torus.

Such treatment is supposed to be presented in one of the forthcoming
publications.

In the aspect of the applications of our geometric material we follow the
general ideology of A.Yu.Morozov, formulated in [15], which proposed to
consider the string theory not as a partial physical theory describing some
narrow class of real phenomena but as one of the universal theories, which
final and, it seems, rather wide area of applications is not understood
completely now.
It is possible that some new applications of certain aspects of
the string theory (and of an infinite dimensional geometry related to it)
will be found during the performing of an interdisciplinary "toy--program" of
an investigation of peculiarities of human vision (in particular, of color
perception) in artificial interactive computerographical systems,
formulated by the author in [16], the results of which may be important for
understanding of various processes in natural interactive (sensorial and
visual) systems.
Hence, our presentation is maximally mathematical and avoids many technical
aspects, which are crucial for some applications; instead of that we prefer
to specially point out a general mathematical meaning of our constructions.

The author thanks Prof.J.-L.Gervais and the collective of Laboratoire de
Phy-\linebreak sique Th\'eorique de l'\'Ecole Normale
Sup\'erieure (Paris) for a friendly imaginative, creative
atmosphere and hospitality, without which the writing of this paper will be
impossible.
The author also specially thanks Prof.J.-L.Gervais, whose ideas on
differential W-geometry provoked him to realize a plan of the article.
Also the author is undebtful to all participants of the seminar on quantum
field theory (especially, Prof.A.Yu.Morozov and A.O.Radul) at the Institute
for Theoretical and Experimental Physics (Moscow), on which the problems
related to W-symmetries were discussed for many times.
The author thanks M.I.Golenishcheva-Kutuzova, B.A.Khesin and V.Yu.Ovsienko,
who being participants of the seminar on representation theory at
Department of Mathematics of Moscow State University introduced the author
into the theme, Prof.M.V.Saveliev (Institute of High Energy Physics,
Serpukhov) for some conversations on differential $W$-geometry at LPTENS,
Prof.L.V.Sabinin and P.O.Mikheev (Dpt.Mathematics, Univ.Peoples' Friendship,
Moscow) for discussions on Lie quasi(pseudo)algebras, Lie
quasi(pseu-\linebreak do)groups and related topics.

\

\head 1. Symplectomorphisms (area-preserving transformations) of torus,
differential operators on a circle and $W$-algebras \endhead

\subhead 1.1. Groups $\Symp(\tor)$, $\Symp^e(\tor)$ and $\Symp_0(\tor)$ of
symplectomorphisms of a torus $\tor$ and their Lie algebras $\Ham(\tor)$
and $\Ham_0(\tor)$ of Hamiltonian and strictly Hamiltonian vector field on
a torus.
Poisson algebras $\Cal F^{\Bbb C}(\tor)$, $\Cal F^{\Bbb C}(T^*(S^1))$ and
$\Cal F^{\Bbb C}(\plane)$ of functions on tori, cylinder and plane,
their deformations and central extensions \endsubhead

Let $(\tor,\omega)$ be a two-dimensional torus with a fixed 2-form (a volume
form) $\omega$ on it.
Let $\Symp(\tor,\omega)$ be {\it a group of symplectomorphisms of a torus\/}
$(\tor,\omega)$, i.e. all diffeomorphisms $\zeta$ of $\tor$ such that
$\zeta_*\omega=\omega$ [17].
So symplectomorphisms of a torus are just {\it its area-preserving
transformations\/}.

It may be easily shown that all groups $\Symp(\tor,\omega)$ are isomorphic.
Indeed, let's identify a torus $(\tor,\omega)$ with the quotient
$\plane/\Gamma$, where the plane $\plane$ with a fixed coordinate system
$(p,q)$ possesses a canonical 2-form $dp\wedge dq$, $\Gamma$ is a certain free
lattice on $\plane$, i.e. a set $\Bbb Z\bold a+\Bbb Z\bold b$,
where $\bold a$ and $\bold b$ are two independent vectors on the
plane.
The symplectomorphisms of a torus may be identified with the classes
$\mod\Gamma$ of functions $\bold f:\plane\mapsto\plane$, which obey some
conditions.
The expression $\mod\Gamma$ means that two functions $\bold f_1$ and
$\bold f_2$ are called equivalent iff
$\bold f_1(\bold x)-\bold f_2(\bold x)\in\Gamma$ for all $\bold x$
from $\plane$.
It should be mentioned that in the coordinate form a function $\bold f$ is
represented as a pair $(f_p,f_q)$ of functions $f_p=f_p(p,q)$ and
$f_q=f_q(p,q)$.
The conditions, to which a function $\bold f$ should obey, are the next ones:
(1) the periodicity condition: $\bold f(\bold x+m\bold a+
n\bold b)=\bold f(\bold x)+m\tilde{\bold a}+n\tilde{\bold b}$ for all $\bold
x$ from $\plane$, where $\tilde{\bold a}$ and $\tilde{\bold b}$ is a pair of
generators of $\Gamma$, i.e. $\tilde{\bold a}=A(\bold a)$,
$\tilde{\bold b}=A(\bold b)$, $A$ is a certain transformation from
$\Aut(\Gamma)$, which is isomorphic to $\SltZ$; (2) the normalization of
Jacobian condition:
$\frac{\bold D\bold f}{\bold D\bold x}=\Ddet\frac{\partial(f_p,f_q)}
{\partial(p,q)}=1$.
The last determinant (Jacobian) is invariant under all linear transformations
of a plane $\plane$.
Because for arbitrary two free lattices $\Gamma_1$ and $\Gamma_2$ there exists
a linear transformation $\gamma$ of $\plane$ such that
$\Gamma_2=\gamma(\Gamma_1)$, the group of symplectomorphisms of
$\plane/\Gamma_1$ and $\plane/\Gamma_2$ may be identified.
So instead of the notation $\Symp(\tor,\omega)$ for a group of
symplectomorphisms of a torus we shall use the simpler notation $\Symp(\tor)$.

It should be mentioned that the group $\Symp(\tor)$ is not connected.
Let us denote the component of the identity in it by $\Symp^e(\tor)$.
Of course, $\Symp^e(\tor)$ is a normal subgroup in $\Symp(\tor)$ and therefore
we have the following strict sequence:
$$ 0\longrightarrow\Symp^e(\tor)\longrightarrow\Symp(\tor)\longrightarrow
\Symp(\tor)/\Symp^e(\tor)\longrightarrow 0.$$
The quotient $\Symp(\tor)/\Symp^e(\tor)$ is isomorphic to {\it the modular
group\/}
$\SltZ$, so that the functions $\bold f$ corresponded to the elements of
$\Symp^e(\tor)$ obey the periodicity condition in the form $\bold f(\bold
x+m\bold a+n\bold b)=\bold f(\bold x)+m\bold a+n\bold b$.

The group $\Symp^e(\tor)$ admits a normal subgroup $\Symp_0(\tor)$ of such
symp-\linebreak lectomorphisms $\zeta$ that
$\int^{1/2}_{-1/2}\int^{1/2}_{-1/2} \bold f(\bold x +t_1\bold a+
t_2\bold b)\,dt_1dt_2=\bold x$. The quotient\linebreak
$\Symp^e(\tor)/\Symp_0(\tor)$ is isomorphic to 2-dimensional compact abelian
group $\tor$, so that a strict sequence
$$ 0\longrightarrow\Symp_0(\tor)\longrightarrow\Symp^e(\tor)\longrightarrow
\tor\longrightarrow 0$$
exists. This strict sequence may be resolved, the resolving mapping
$\tor\to\Symp^e(\tor)$ realises a group $\tor$ as a group of
movements of a torus $\tor$.

The Lie algebra of the group $\Symp^e(\tor)$ is {\it an algebra of
Hamiltonian vector
fields on a torus\/}, i.e. the fields $\xi$ such that $\Cal L_\xi\omega=0$.
It should be mentioned that after an identification of a torus $\tor$ with the
quotient $\plane/\Gamma$ Hamiltonian fields $\xi$ on a torus may be also
characterized as divergence free ones, i.e. vector fields $\xi$ such that
$\Ddiv\xi=0$.
{}From Cartan theorem it follows that $\xi$ is Hamiltonian iff $d\alpha_\xi=0$,
where $\alpha_\xi=\imath_\xi\omega$.
So the algebra $\Ham(\tor)$ of Hamiltonian vector fields on a torus may be
identified with an algebra of closed 1-forms on a torus with respect to the
bracket $\{\alpha,\beta\}=d\frac{\alpha\wedge\beta}{\omega}$.

The Lie algebra of the group $\Symp_0(\tor)$ is a certain subalgebra
of $\Symp^e(\tor)$, {\it the algebra of strictly Hamiltonian vector fields\/},
i.e. the Hamiltonian vector fields $\xi$ such that the corresponding 1-form
$\alpha_\xi$ is strict.
The Lie algebra $\Ham_0(\tor)$ of strictly Hamiltonian vector fields on a
torus is an ideal in the algebra $\Ham(\tor)$ so that a strict sequence
$$ 0\longrightarrow\Ham_0(\tor)\longrightarrow\Ham(\tor)\longrightarrow
\plane\longrightarrow 0$$
exists.
The quotient $\plane=\Ham(\tor)/\Ham_0(\tor)$ may be interpreted as the
cohomology group $H^1(\tor,\Bbb R)$ if one considers its as a quotient of a
space of closed 1-forms by a subspace of strict ones.
This strict sequence may be resolved.
The resolving mapping $\plane\longmapsto\Ham(\tor)$ realises the vector space
$\plane$ as a certain parallelization of a torus.

It should be mentioned that if 1-form $\alpha_\xi$ is strict then there exist a
function $F_\xi$ on a
torus such that $\alpha_\xi=dF_\xi$.
The bracket in the space of strictly Hamiltonian vector fields induces a
bracket in the space $\Cal F(\tor)$ of functions (0-forms) on a torus.
The bracket in the space $\Cal F(\tor)$ has the form $\{F,G\}=\frac{dF\wedge
dG}\omega$.
The center of the Lie algebra $\Cal F(\tor)$ consists of constant functions,
so it may be identified with 1-dimensional space $\Bbb R$.
The Lie algebra of strictly Hamiltonian vector fields on a torus is isomorphic
to a quotient $\Cal F(\tor)/\Bbb R$.
The isomorphism is realised by an exterior derivative $d$, which maps the above
quotient onto the Lie algebra of strict 1-forms, which is isomorphic to
$\Ham_0(\tor)$.
The mapping from the algebra $\Cal F(\tor)$ to the algebra $\Ham_0(\tor)$ is
symbolized as $\sgrad$ and is called by {\it "skew-gradient",} so that
$\xi=\sgrad(F_\xi)$ for all strictly Hamiltonian vector fields on a torus.
It should be mentioned that a strict sequence
$$ 0\longmapsto\Bbb R\longmapsto\Cal F(\tor)\longmapsto\Ham_0(\tor)\longmapsto
0$$
may be resolved. The resolving mapping identify the algebra $\Ham_0(\tor)$
with the subspace $\Cal F_0(\tor)$ of functions of zero mean value.

It should be mention that the bracket $\{\cdot,\cdot\}$ on the space $\Cal
F(\tor)$ of functions on a torus, determines a structure of a Poisson algebra
on it.
This algebra is called by {\it the Poisson algebra of functions on a torus\/}.
It means that $\{F,GH\}=\{F,G\}H+G\{F,H\}$ for all functions $F,G,H$ from $\Cal
F(\tor)$.
In another words the bracket $\{\cdot,\cdot\}$ is a derivation of the ordinary
commutative multiplication of functions.

Let us realise torus $\tor$ as a quotient $\plane/\Gamma$, where
$\Gamma=\{(mM,nN)\in\plane, m,n\in\Bbb Z; M,N$ are fixed real positive
numbers$\}$.
Then there is defined a canonical basis in $\Cal
F^{\Bbb C}(\tor)$, the complexification of the algebra $\Cal F(\tor)$:
$e_{m,n}=\Cal D(M,N)\exp(\frac{imp}M)\mathbreak\exp(\frac{inq}N)$, where
it is convenient to put a normalization constant $\Cal D(M,N)$ being equal to
$MN$.
The Poisson brackets $\{\cdot,\cdot\}$ in $\Cal F(\tor)$ or $\Cal F^{\Bbb
C}(\tor)$ have the form $\{F,G\}=\frac{\partial F}
{\partial p}\frac{\partial G}{\partial q} - \frac{\partial G}{\partial p}
\frac{\partial F}{\partial q}$ so that [17]
$$ \{e_{m,n},e_{m',n'}\}=(mn'-nm')e_{m+m',n+n'}. $$
The subalgebra $\Cal F_0^{\Bbb C}(\tor)$ of $\Cal F^{\Bbb C}(\tor)$ (the
complexification of $\Cal F_0(\tor)$) is realized as one spanned by generators
$e_{m,n},(m,n)\ne(0,0)$.

Let us call the algebra spanned formally by $e_{m,n}$ by {\it
Floratos-Iliopoulos
algebra\/} (see [18], where it was considered).
It should be mentioned that generators $L_k, k\in\Bbb Z$ of the algebra
$\CVect(\circle)$ of vector fields on a circle $\circle$, obeying the
commutation relations $[L_i,L_j]=(i-j)L_{i+j}$ may be expressed
formally via $e_{m,n}$ in the following way [18]:
$$ L_k=\sum_{m\in\Bbb Z} \Cal G_k(m)e_{m,k},$$
where $\Cal G_k(m)=\frac{(-1)^m}m$ if $m\ne 0$ and $0$ otherwise.
Of course, under the verification of the commutations relations of generators
$L_k$ some rather natural regularization is used.
So the Lie algebra $\CVect(\circle)$ (more precisely, so called {\it Witt
algebra,}
which is spanned by generators $L_k$ of $\CVect(\circle)$, but we
shall not mention a difference between them below) is realized as a certain
subalgebra of the Floratos-Iliopoulos algebra.

Under the action of the algebra $\CVect(S^1)$ the Floratos-Iliopoulos algebra
may be decomposed into the sum of tensor modules of positive integer weights,
the corresponding tensor operators will be denoted by $w^{(i)}_n$ ($i\ge
2$).
It should be mentioned that $L_n=w^{(2)}_n$, the commutation relations between
generators $w^{(i)}_n$ have the form
$$ [w^{(i)}_n,w^{(j)}_m]=((j-1)n-(i-1)m)w^{(i+j-2)}_{n+m}, $$
$w^{(i)}_n$ themselves are defined as
$$ w^{(i)}_k=\sum_{k\in\Bbb Z}\Cal G^{(i)}_k(m)e_{k,m}, $$
$\Cal G^{(i)}_k(m)=P_i(\frac1m)$, where the polynomials $P_i(x)$ of degree
$i-1$ are connected by the easily calculated requrrent formulas.

The algebra $\Ham_0(\tor)$ of strictly Hamiltonian vector fields on a torus
admits a central extension ${\widehat\Ham}_0(\tor)$, which may be defined in a
following way $[\xi,\eta]_\partial=[\xi,\eta]+\int_{\tor}[\partial,\xi]\eta\,
dpdq$, where $\partial$ is a certain Hamiltonian but not strictly Hamiltonian
vector field on the torus.
One may choose $\partial=a\partial_p+b\partial_q$ and this is a general
setting up to trivial central extensions.
So the universal central extension of $\Ham_0(\tor)$ is defined by the
strict sequence
$$ 0\longrightarrow H_1(\tor,\Bbb R)\longrightarrow{\widehat\Ham}_0(\tor)
\longrightarrow\Ham_0(\tor)\longrightarrow 0, $$
where $H_1(\tor,\Bbb R)$ is a 2-dimensional group of homology of
torus with real coefficients.

The central extension ${\widehat\Ham}_0(\tor)$ of the Lie algebra
$\Ham_0(\tor)$ induces a central extension ${\widehat{\Cal F}}^{\Bbb C}(\tor)$
of the algebra $\Cal F^{\Bbb C}(\tor)$, defined as $\{F,G\}_\partial=
\{F,G\}+\int_{\tor}\Cal L_\partial F\cdot G\, dpdq$, so that the following
strict sequence
$$ 0\longrightarrow\Bbb C^2\longrightarrow{\widehat{\Cal F}}^{\Bbb C}(\tor)
\longrightarrow\Cal F^{\Bbb C}(\tor)\longrightarrow 0$$
exists.
This central extension has the following form in the basis $e_{m,n}$:
$$ [e_{m,n},e_{m',n'}]=(mn'-nm')e_{m+m',n+n'}+(am+bn)\delta(m+m')\delta(n+n')
\boldkey 1$$

The extended Floratos-Iliopoulos algebra admits an embedding of the Virasoro
algebra $\Cvir$ with a non-trivial central charge into it (one of possible
embeddings was described in 18]).

The Poisson algebra $\Cal F^{\Bbb C}(\tor)$ may be deformed into a certain
associative algebra [19].
In the canonical basis such deformation has the form
$$ e_{m,n}\cdot e_{m',n'}=\exp(i\lambda(mn'-nm'))e_{m+m',n+n'}. $$
The corresponding commutator algebra has the form (after a renormalization of
generators)
$$ [e_{m,n},e_{m',n'}]=\frac1\lambda\sin(\lambda(nm'-mn'))e_{m+m',n+n'}.$$
So there is defined a certain deformation of Floratos-Iliopoulos algebra,
which is called (centreless) {\it Sine-algebra (or Fairlie-Fletcher-Zachos
algebra)} [20].
The corresponding deformation of $\Ham_0(\tor)$ is called {\it the Lie algebra
of
strictly Hamiltonian vector fields on a quantum torus\/} and is denoted by
$\Ham_0(\tor_q)$($q=e^{i\lambda}$).

(Centreless) Sine algebra admits a central extension, which has the form
$$ [e_{m,n},e_{m',n'}]=\frac1\lambda\sin(\lambda(nm'-mn'))e_{m+m',n+n'}+
(am+bn)\delta(m+m')\delta(n+n')\boldkey 1.$$
The Sine-algebra with a center also admits the embedding of the Virasoro
algebra $\Cvir$:
$$ L_k=\sum_{m\in\Bbb Z}\widetilde{\Cal G}_k(m)e_{k,m}, $$
where $\widetilde{\Cal G}_k(m)$ are defined by formulas, analogous to one of
[18].

Under the action of the algebra $\Cvir$ Sine-algebra may be also decomposed
into the sum of tensor modules of positive integer weights, the corresponding
tensor operators will be denoted by $\tilde w^{(i)}_n$ ($i\ge 2$).
Their commutation relations have the form (for the centreless Sine-algebra)
$$ [\omega^{(k)}_n,\omega^{(l)}_m]=\sum_{p\ge 0} A^p_{kl}
\omega^{(k+l-2p-1)}_{m+n},$$
where $A^p_{kl}=\sum_{i+j=2p+1}(-1)^iC^i_{k-1}C^j_{l-1}$,
$C^m_n=\frac{n!}{m!\,(n-m)!}$ if $m\le n$ and $0$ otherwise.

Now let us mention that all our constructions admit certain limits with one or
both parameters $N$ and $M$ tends to infinity.
That means that the torus becomes a cylinder or a plane.
In the case $M\to\infty$ the Poisson algebra $\Cal F^{\Bbb C}(\tor)$
inverts into {\it the Poisson algebra $\Cal F^{\Bbb C}(T^*\circle)$
of functions on a cylinder.}
There is a natural basis in $\Cal F^{\Bbb C}(T^*\circle)$:
$ e^m_n=-i\exp(\frac{inq}N)p^m$, in which the Poisson brackets have the
form:
$$ \{e^m_n,e^{m'}_{n'}\}=(mn'-nm')e^{m+m'-1}_{n+n'}.$$
The embedding of the algebra $\CVect(\circle)$ into $\Cal F^{\Bbb
C}(T^*\circle)$ is natural: $L_k\mapsto e^1_k$.
Such embedding admits, indeed an one-parametric deformation $L_k\mapsto
e^1_k+(k+1)\lambda e^0_k$, $\lambda$ is a parameter.
The central extension $\widehat{\Cal F}^{\Bbb C}(T^*\circle)$ of $\Cal F^{\Bbb
C}(T^*\circle)$ has the form:
$$
\{e^m_n,e^{m'}_{n'}\}=(mn'-nm')e^{m+m'-1}_{n+n'}+
\frac1{12}(n^3-n)\delta(n+n')\delta_{m,1}\delta_{m',1}\boldkey 1$$
The embeddings of $\CVect(\circle)$ into $\Cal F^{\Bbb C}(T^*\circle)$ extends
to embeddings of the Virasoro algebra $\Cvir$ into $\widehat{\Cal F}^{\Bbb
C}(T^*\circle)$.

The Poisson algebra $\Cal F^{\Bbb C}(T^*\circle)$ is deformed into
{\it the algebra $\DOP^{\Bbb C}(\circle)$ of differential operators on
a circle.}
The corresponding commutator algebra is {\it a Lie algebra
$\LDOP^{\Bbb C}(\circle)$ of differential operators on a circle.}
The embeddings of $\CVect(\circle)$ in $\Cal F^{\Bbb C}(T^*\circle)$ induce
embeddings of this algebra into $\LDOP^{\Bbb C}(\circle)$; these embedding
are the same if one consider a subalgebra $\LDOP^{\Bbb C}(\circle)_+$ of
$\LDOP^{\Bbb C}(\circle)$, the commutator algebra of the algebra $\DOP^{\Bbb
C}(\circle)_+$ of differential operators on the circle $\circle$ without free
terms.
Both algebras $\LDOP^{\Bbb C}(\circle)$ and $\LDOP^{\Bbb C}(\circle)_+$ admit
central extensions $\cLDOP^{\Bbb C}(\circle)$ and $\cLDOP^{\Bbb C}(\circle)_+$,
which are defined by {\it the Kac-Peterson cocycle\/}
$c(f(q)D^m,g(q)D^n)=\frac{m!\,n!}{(m+n+1)!}\int_{\circle}f^{(n)}(q)g^{(m+1)}(q)
\, dq$, where $D=\partial/\partial q$, $f^{(n)}(q)=\frac{\partial^n f(q)}
{\partial q^n}$ [21](see also [6]).
The embeddings of $\CVect(\circle)$ in $\LDOP^{\Bbb C}(\circle)$ (resp. the
emdedding in $\LDOP^{\Bbb C}(\circle)_+$) are extended to embeddings of the
Virasoro algebra $\Cvir$ in $\cLDOP^{\Bbb C}(\circle)$ (resp. an embedding in
$\cLDOP^{\Bbb C}(\circle)_+$).

To make the construction of central extensions $\cLDOP^{\Bbb C}(\circle)$ and
$\cLDOP^{\Bbb C}(\circle)_+$ of $\LDOP^{\Bbb C}(\circle)$ and $\LDOP^{\Bbb
C}(\circle)_+$ more clear one should transit to {\it the algebra $\PDOP^{\Bbb
C}(\circle)$ of all pseudodifferential operators on a circle\/} $\circle$
[5-8].
The commutator algebra $\LPDOP^{\Bbb C}(\circle)$ of this algebra,
{\it the Lie algebra of pseudodifferential operators on a circle\/},
admits a central extension, which may be defined as
$[A,B]_c=[A,B]+c\cdot\Tr([A,\log D],B])$, where
$\Tr(A)=\int\Res(A)\,dq$, $\Res(A)=a_{-1}(q)$ if $A=\sum_{k\in\Bbb Z}a_k(q)D^k$
[5].
The cocycle $c(A,B)=\Tr([A,\log D],B)$ is so-called {\it Kravchenko-Khesin
cocycle\/}.
The extension by Kravchenko-Khesin cocycle being restricted on subalgebras
$\LDOP^{\Bbb C}(\circle)$ and $\LDOP^{\Bbb C}(\circle)_+$ define extensions
described above.

In the case $M,N\to\infty$ the Poisson algebra $\Cal F^{\Bbb C}(\tor)$ inverts
into the Poisson algebra $\Cal F^{\Bbb C}(\plane)$ of functions on a plane
$\plane$.
There is defined a natural polynomial basis in $\Cal F^{\Bbb C}(\plane)$:
$e^{mn}=p^mq^n$, in which the Poisson brackets have the form
$$ \{e^{mn},e^{m'n'}\}=(mn'-nm')e^{m+m'-1,n+n'-1}.$$

The Poisson algebra $\Cal F^{\Bbb C}(\plane)$ may be deformed into the Weyl
algebra $W_{\Bbb C}(\plane)$.
Polynomial elements of this algebra may be represented by differential
operators on a line $\Bbb R^1$, so there is corresponded a certain
differential operator $\Cal D_P$ to the polynomial $P$ from $\Cal F^{\Bbb
C}(\plane)$, where $P$ is a Weyl (symmetric) symbol of $\Cal D_P$.
The commutator algebra of the Weyl algebra $W_{\Bbb C}(\plane)$ is so-called
Moyal algebra $\Moy_{\Bbb C}(\plane)$.
The commutator in the Moyal algebra has the form [19,22]
$$ [F,G]=\sum_{p\in\Bbb Z}(-1)^p\sum_{i+j=2p+1}\frac{(-1)^i}{i!\,j!}
(\partial^i_p\partial^j_qF)(\partial^j_p\partial^iG)$$
or in the basis $e^{mn}$
$$ [e^{mn},e^{m'n'}]=\sum_{p\ge 0}(-1)^pA_{mm'nn'}^{2p+1}
e^{m+m'-2p-1,n+n'-2p-1},$$
where $A_{mm'nn'}^p=\sum_{i+j=p}(-1)^ii!\,j!\,C^i_nC^j_nC^j_{m'}C^i_{n'}$,
$C^m_n=\frac{n!}{m!\,(n-m)!}$ if $m\le n$ and $0$ otherwise.
\pagebreak

\subhead 1.2. Poisson algebra $\Cal F^{\Bbb C}(T^*\circle)$, its central
extension $\widehat{\Cal F}^{\Bbb C}(T^*\circle)$ and $w_\infty$-algebra
(Bakas algebra); Lie algebra $\LDOP^{\Bbb C}(\circle)$, its central extension
$\cLDOP^{\Bbb C}(\circle)$ and $W_\infty$-algebras (Pope-Shen-Romans
algebras).
Gelfand-Dickey algebras $\GD^{\Bbb C}_n$ and Radul mapping $\Cal F(\LDOP^{\Bbb
C}(\circle)^*)\mapsto\GD^{\Bbb C}_n$; Radul bundle $\Rad^{\Bbb C}(\Cal M_n)$
and the Lie algebra of its sections $\Rad_{\Bbb C}$ (Radul algebra).
Wedge subalgebra of Pope-Sheen-Romans $W_\infty$-algebra; associative model
algebras $\Md^{(\lambda)}(\sltC)$ for the Lie algebra $\sltC$ and their
commutator algebras $\LMd^{(\lambda)}(\sltC)$ - model Lie algebras (or Feigin
algebras) for $\sltC$; Racah-Wigner algebras $\RW^{(\lambda)}_\infty(\sltC)$
for $\sltC$, their reductions $\RW^{(\lambda)}_n(\sltC)$ of order $n$ and
central extensions $\cRW^{(\lambda)}_n(\sltC)$ of $\RW^{(\lambda)}_n(\sltC)$
\endsubhead

The Poisson algebra $\Cal F^{\Bbb C}(\circle)$ may be expanded under the
action of $\CVect(\circle)$ into the sum of tensor modules of positive integer
weights, the corresponding tensor operators $w^{(i)}_k$ ($i\ge 1$) are equal
to $e^{i-1}_k$ so that the commutation relations
$$ [w^{(i)}_m,w^{(j)}_n]=((j-1)n-(i-1)m)w^{(i+j-2)}_{n+m}$$
holds.
The algebra spanned by $w^{(i)}_n$, $n\in\Bbb Z$, $i\ge 2$ is closed and is
called {\it $w_\infty$-algebra\/} (or {\it Bakas algebra\/}) [23] as well as
its
central extension, which is obtained from $\widehat{\Cal F}^{\Bbb
C}(T^*\circle)$.

Now we are interested in the deformations of $w_\infty$-algebra, when
$\widehat{\Cal F}^{\Bbb C}(T^*\circle)$ deforms into $\cLDOP^{\Bbb
C}(\circle)$.
Such deformations were found by C.N.Pope, X.Shen and L.J.Romans [24] (see also
[25]), so the
corresponding algebras are called {\it Pope-Shen-Romans-algebras\/} (or {\it
$W_\infty$-algebras\/}).
The explicit formulas for their generators were found by I.Bakas, B.Khesin and
E.Kiritsis [8] so the embedding of $W_\infty$-algebras in $\cLDOP^{\Bbb
C}(\circle)$ are called {\it Bakas-Khesin-Kiritsis embeddings\/}.

Let us at first construct so-called $W_{1+\infty}$-algebra,
which is an algebra of $\Cvir$-tensor operators in $\cLDOP^{\Bbb C}(\circle)$,
following [8].
Its generators $V^s_m$ are labelled by $m\in\Bbb Z$ and $s\ge 1$.
Their explicit form is
$$ V^s_m=-B(s)\sum^s_{k=1}\alpha^s_kC^{m+s-1}_{k-1}z^{m+s-k}D_z^{s-k},$$
where $B(s)=\frac{2^{s-3}(s-1)!}{(2s-3)!!}$,
$\alpha^s_k=\frac{(2s-k-1)!}{[(s-k)!\,]^2}$;
$z=\exp(iq)$, $D_z=\frac{\partial}{\partial z}$.
To describe commutation relations in $W_{1+\infty}$-algebra it is convenient
to introduce some notations
$$\align
g^{ss'}_l(m,n;\mu):= & {\ssize\frac1{2(l+1)!}}f^{ss'}_l(\mu)N^{ss'}_l(m,n) \\
f^{ss'}_l(\mu):= & \ssize\sum_{k\ge 0}\frac{(-\frac12-2\mu)_k(\frac32+2\mu)_k
(-\frac{l+1}2)_k(-\frac
l2)_k}{k!\,(-s+\frac32)_k(-s'+\frac32)_k(s+s'-l-\frac32)_k} \\
N^{ss'}_l(m,n):= &\ssize
\sum^{l+1}_{k=0}{(-1)^kC^{l+1}_k(2s-l-2)_k[2s'-k-2]_{l+1-k}[s-1+m]_{l+1-k}\*[s'-1+n]_k}
\endalign
$$
where $(a)_k:=a(a+1)\ldots(a+k-1)$, $[a]_k:=a(a-1)\ldots(a-k+1)$.
Then
$$\align
[V^s_m,V^{s'}_n]=&((s'-1)m-(s-1)n)V^{s+s'-2}_{m+n}+c_s(m,\mu)\delta_{ss'}\delta(m+n)+
\\
&\sum_{r\ge 0}g^{ss'}_{2r}(m,n;\mu)V^{s+s'-2r-2}_{m+n},\endalign $$
where $\mu=-\frac12$,
$$
c_s(m;-\frac12)=c\cdot\frac{(m+s-1)!}{(m-s)!}\cdot
\frac{2^{2(s-3)}[(s-1)!\,]^2}{(2s-1)!!\,(2s-3)!!}.$$

$W_{1+\infty}$-algebra is not a deformation of Bakas algebra because it
contains tensor operators of upper index 1.
To construct the correct deformation (which is the "right" $W_\infty$-algebra)
one
should to transform generators $V^s_m$ in the following manner.
Let us introduce new generators $W^s_m$ ($m\in\Bbb Z$, $s\ge 2$) by the
formulas
$$
W^s_m=V^s_m+\frac{B(s)}{s-1}\sum_{k=1}^{s-1}(-1)^l\frac{(2s-2l-1)}{B(s-l)}\cdot\frac{(m+s-1)!}{(m+s-l-1)!}V^{s-1}_m$$
or explicitely
$$
W^s_m=-\frac{B(s)}{s-1}\sum^{s-1}_{k=1}\beta^s_kC^{m+s-1}_{k-1}z^{m+s-k}D_z^{s-k},$$
where $\beta^s_k=\frac{(2s-k-1)!}{(s-k)!\,(s-k-1)!}$, i.e. $W_\infty$-algebra
is realized in $\LDOP^{\Bbb C}(\circle)_+$.
The commutation relations in $W_\infty$-algebra have the form
$$\align
[W^s_m,W^{s'}_n]=&((s'-1)m-(s-1)n)W^{s+s'-2}_{m+n}+c_s(m;\mu)\delta_{ss'}\delta(m+n)+
\\&\sum_{r\ge 1}g^{ss'}_{2r}(m,n;\mu)W^{s+s'-2-2r}_{m+n},\endalign$$
where $\mu=0$,
$$ c_s(m,0)=\frac c2\cdot\frac{(m+s-1)!}{(m-s)!}\cdot\frac{2^{2(s-3)}s!\,
(s-2)!}{(2s-1)!!\,(2s-3)!!}.$$

Let us consider following [6,7] an affine subspace $\Cal M_n$ in the algebra
$\DOP^{\Bbb C}(\circle)$ of the form $\Cal M_n=\{A\in\DOP^{\Bbb C}(\circle)$:
$A=D^n+a_{n-1}(q)D^{n-1}+\ldots+a_1(q)D+a_0(q)\}$.
The tangent space $T_A(\Cal M_n)$ of manifold $\Cal M_n$ at the point $A$ may
be
identified with the space of all operators $X$ from $\DOP^{\Bbb C}(\circle)$
of the form $X=x_{n-1}(q)D^{n-1}+\ldots+x_0(q)$.
The corresponding cotangent space $T^*_A(\Cal M_n)$ may be identified with the
space of all operators $Y$ from $\PDOP^{\Bbb C}(\circle)$ of the form
$Y=y_{-1}(q)D^{-1}+\ldots+y_{1-n}(q)D^{1-n}+y_{-n}(q)D^{-n}$.
The spaces $T^*_A(\Cal M_n)$ and $T_A(\Cal M_n)$ are paired by
$<X,Y>=\Tr(XY)$.

There is defined [6,7] a tensor operator field $V:T^*(\Cal M_n)\mapsto T(\Cal
M_n)$, namely $V_A$, which maps $T^*_A(\Cal M_n)$ to $T_A(\Cal M_n)$, is
defined by the following formulas $V_A(Y)=A(YA)_+-(AY)_+A$, $Y\in T^*_A(\Cal
M_n)$, where $A\mapsto A_+$ is the natural projection of $\PDOP^{\Bbb
C}(\circle)$ to $\DOP^{\Bbb C}(\circle)$.

The space $\Cal F(\Cal M_n)$ of functions on the manifold $\Cal M_n$ possesses
a structure of the Poisson algebra via $\{F,G\}=<V(dF),dG>$ [6,7].
This Poisson algebra is called {\it Gelfand-Dickey algebra\/} and is denoted
by $\GD_n^{\Bbb C}$.
The center of Gelfand-Dickey algebra $\GD_n^{\Bbb C}$ consists of constant
functions, so the strict sequence
$$0\longrightarrow\Bbb C\longrightarrow\GD_n^{\Bbb
C}\longrightarrow\overline{\GD}^{\Bbb C}_n\longrightarrow 0$$
exists, where $\overline{\GD}^{\Bbb C}_n$ is a quotient $\GD^{\Bbb C}_n/\Bbb
C$.

Let us now construct {\it the Radul mapping\/} from the Poisson algebra
\linebreak
$\Cal F(\LDOP^{\Bbb C}(\circle)^*)$ of the functions on the coadjoint
module $\LDOP^{\Bbb C}(\circle)^*)$ of the Lie algebra $\LDOP^{\Bbb
C}(\circle)$ onto the Poisson algebra $\GD_n^{\Bbb C}$.
Namely, let us construct a mapping $H:\LDOP^{\Bbb C}(\circle)\mapsto
\CVect(\Cal M_n)$, where the mapping $H_A:\LDOP^{\Bbb C}(\circle)\mapsto
T_A(\Cal M_n)$ is defined by the formulas $H_A(B)=V_{(BA^{-1})_-}(A)=
AB-(ABA^{-1})_+A$; $B\in\LDOP^{\Bbb C}(\circle)$, $A_-=A-A_+$ [6,7].
It should be mentioned that $\Ker(H_A)=\{C\in\LDOP^{\Bbb C}(\circle)$: $C=BA$,
$A\in\DOP^{\Bbb C}(\circle)\}$.
It is stated that $\Iim(H)$ is contained in $\Iim(d)$, where
$d:\GD^{\Bbb C}_n\mapsto \CVect(\Cal M_n)$.
Because $\Ker(d)$ consists of constant functions on $\Cal M_n$ then $H$
realizes a homomorphism of $\LDOP^{\Bbb C}(\circle)$ into
$\overline{\GD}_n^{\Bbb C}$ considered as a Lie algebra.
Such homomorphism is extended to the homomorphism of the Poisson algebra
$\Cal F(\LDOP^{\Bbb C}(\circle)^*)$ onto the Gelfand-Dickey algebra
$\GD_n^{\Bbb
C}$.
Of course, one may consider a restriction of the Radul mapping $\Cal
F(\LDOP^{\Bbb C}(\circle)^*)\mapsto\GD^{\Bbb C}_n$ on the Poisson subalgebra
$\Cal F(\LDOP^{\Bbb C}(\circle)_+^*)$ of the Poisson algebra $\Cal
F(\DOP^{\Bbb C}(\circle)^*)$, where $\DOP^{\Bbb C}(\circle)_+$ is a
subalgebra of $\DOP^{\Bbb C}(\circle)$ of differential operators without free
terms, $\LDOP^{\Bbb C}(\circle)_+$ is its commutator algebra, the Lie algebra
of differential operators without free terms, and $\LDOP^{\Bbb
C}(\circle)^*_+$
is the coadjoint module over this Lie algebra.

Now let us construct so-called {\it Radul bundle\/} and the Lie algebra of its
sections ({\it Radul algebra\/}).
Let us consider a trivial bundle $\Rad^{\Bbb C}(\Cal M_n)$ over $\Cal M_n$ with
fiber
isomorphic to $\LDOP^{\Bbb C}(\circle)$.
One may define the following commutator in the space of its sections:
if $E,F\in\Cal O(\Rad^{\Bbb C}(\Cal M_n))$ then
$$\align
[E,F]'(A):=&[E(A),F(A)]+V_{(EA^{-1})_-}(F)-V_{(FA^{-1})_-}(E)\in\Rad_A^{\Bbb
C}(\Cal
M_n),\\ &\text{where}\quad V_X(E(A))=\left.\frac{\partial}{\partial
t}E(A+tV_X(A))\right|_{t=0},\quad A\in\Cal M_n.\endalign$$
The algebra of sections
of the bundle $\Rad^{\Bbb C}(\Cal M_n)$ will be called the Radul algebra
$\Rad_{\Bbb C}$.
As it was stated in [7] the mappings $H_A$ being glued define a
homomorphism of the Radul algebra $\Rad_{\Bbb C}$ onto $\overline{\GD}^{\Bbb
C}_n$ considered as a Lie algebra.

The algebra $\GD_n^{\Bbb C}$ defined above is just Gelfand-Dickey algebra
$\GD(\gl(n,\Bbb C))$
for the Lie algebra $\gl(n,\Bbb C)$. It is more convenient to consider
Gelfand-Dickey
algebra $\GD(\ssl(n,\Bbb C))$ for the Lie algebra $\ssl(n,\Bbb C)$. To do it
one should
consider the subspace $\Cal M_n^{(0)}$ of $\Cal M_n$ of operators $A$ with
$a_{n-1}=0$, it should be mentioned that our presentation may be considered
for $\GD(\ssl(n,\Bbb C))$ if one change $\LDOP^{\Bbb C}(\circle)$ on
$\LDOP^{\Bbb
C}(\circle)_+$.

As it was marked in [24] generators $W^s_m$ of the Pope-Shen-Romans
$W_\infty$-algebra with $1-s\le m\le s-1$ form a closed Lie algebra, which is
called {\it wedge subalgebra\/} of $W_\infty$-algebra.
The generators $W^2_{-1}$, $W^2_0$, $W^2_1$ of the wedge subalgebra are just
the generators of the Lie algebra $\sltC$.
A space of the wedge subalgebra as $\sltC$-module is identified with the
model $M_{\oddim}(\sltC)$ of odd-dimensional representations of the Lie algebra
$\sltC$ [26].

Let us analyse the algebraic structures related to the model
$M_{\oddim}(\sltC)$
more systematically.
If one considerd the universal envelopping algebra $\Cal U(\sltC)$ of the Lie
algebra $\sltC$ and the ideal $I_\lambda$, generated by $K-\lambda$, where $K$
is a Casimir operator for $\sltC$ (i.e. a quadratic element of the center
$\Cal Z(\Cal U(\sltC))$ of the universal envelopping algebra $\Cal U(\sltC)$),
in it, then the quotient $\Cal U(\sltC)/I_\lambda$ supplies the model
$M_{\oddim}(\sltC)$ by a structure of an associative algebra.
Such associative algebra will be called {\it the associative model algebra for
the Lie algebra $\sltC$\/} and will be denoted by $\Md^{(\lambda)}(\sltC)$.

The commutator algebra $\LMd^{(\lambda)}(\sltC)$ of the associative model
algebra
\linebreak $\Md^{(\lambda)}(\sltC)$ will be called {\it the model Lie algebra
for the Lie
algebra $\sltC$\/} (or {\it Feigin algebra\/}, see [27]).
The wedge subalgebra of the Pope-Shen-Romans $W_\infty$-algebra is just the
Feigin algebra $\LMd^{(0)}(\sltC)$.

Let us now consider the Racah-Wigner algebras, their reductions of finite
order and central extensions of such reductions for the Lie algebra $\sltC$.
{\it The Racah-Wigner algebra $\RW^{(\lambda)}_\infty(\sltC)$ for the Lie
algebra
$\sltC$\/} is the universal envelopping algebra $\Cal
U(\LMd^{(\lambda)}(\sltC))$ of the model Lie algebra
$\LMd^{(\lambda)}(\sltC)$.

\definition{Definition 1} An associative algebra $\RW^{(\lambda)}_n(\sltC)$
is called {\it a reduced Racah-Wigner algebra of order $n$ for the Lie algebra
$\sltC$\/} iff (1) it admits a homomorphism onto $\Md^{(\lambda)}(\sltC)$;
(2) $\RW^{(\lambda)}_n(\sltC)$ is generated by the direct sum
$\pi_1\oplus\pi_2\oplus\ldots\oplus\pi_{n-1}\oplus\pi_n$ of the first $n$
odd-dimensional representations of $\sltC$ (this direct sum is isomorphic to
$\slnC$ as $\sltC$-module), the natural action of $\sltC$ is defined by
$\ad(\pi_1)$; (3) $\RW^{(\lambda)}_n(\sltC)$
as $\sltC$-module is isomorphic to $S^{\cdot}(\slnC)$
($S^{\cdot}(V)$ is the symmetric algebra over $V$).
\enddefinition

Reduced Racah-Wigner algebras $\RW^{(\lambda)}_2(\sltC)$ of order $2$
with non-homoge-\linebreak neous quadratic relations were considered in [28].
It was shown that
$\RW^{(\lambda_1)}_2(\sltC)\mathbreak \simeq\RW^{(\lambda_2)}_2(\sltC)$ for
arbitrary
$\lambda_1$ and $\lambda_2$.

\proclaim{Proposition 1} There exists a reduced Racah-Wigner algebra
$\RW^{(\lambda)}_n(\sltC)$ of finite order n for the Lie algebra $\sltC$ for
all $n$.
\endproclaim
\demo{Proof} Unfortunately, we do not know the direct algebraic proof of this
proposition, which is, certainly, preferable.
Let us formulate a quantum-field one, which is based on formalism of the
$q_R$-conformal field theory [28].
Namely, let us considered the set of $q_R$-affine currents, which charges form
the Lie algebra $\slnC$ (or which components form $q_R$-affine Lie
algebra) [28].
Let us consider the set of Casimir operators $K_1,\ldots, K_{n-1}$ and the
higher spin fields $W^2(z),\ldots, W^n(z)$, which correspond to them
in the operator algebra of $q_R$-affine currents.
The algebra, generated by the components $W^s_m$ ($1-s\le m\le s-1$;
$W^s(z)=\sum_{m\in\Bbb Z}W^s_m z^{-s-m}$) is just the reduced Racah-Wigner
algebra $\RW^{(\lambda)}_n(\sltC)$ of the order $n$ for the Lie algebra
$\sltC$, where $\lambda=\frac14(q_R^{-1}+3)(q_R^{-1}+1)$.
\enddemo

The non-linear structure constants of the constructed reduced Racah-Wigner
algebras $\RW^{(\lambda)}_n(\sltC)$ admits a deformation by the linear
structure constants of the universal envelopping algebra $\Cal U(\slnC)$
of the Lie algebra $\slnC$.
One may introduce a new central element $\rho$ and consider this deformation
as a central extension of the reduced Racah-Wigner algebra.
Such central extension will be denoted by $\cRW^{(\lambda)}_n(\slnC)$.
If one consider the fixed value of central element $\rho$, i.e. a quotient
of $\cRW^{(\lambda)}_n(\slnC)$ by the ideal generated by $\rho-\rho_0$, then
the limit $\rho_0\to\infty$ inverts the obtained quotient into the universal
envelopping algebra $\Cal U(\slnC)$.
So one may consider the algebras $\cRW^{(\lambda)}_n(\sltC)$ as certain
deformations of $\Cal U(\slnC)$.

\head 2. Gervais-Matsuo differential $W$-geometry and $W$-symmetries of a
second quantized free string \endhead

In this paragraph we shall work with the following objects [1](see also
[13,29-31]):
\roster
\item"$Q$" (or the dual $Q^*$) --- the space of external degrees of a freedom
of
a string.
The coordinates $x^\mu_n$ on $Q$ are the Taylor coefficients of functions
$x^\mu(z)$, which determines a world-sheet of a string in a complexified target
space.
\item"$M(\Vir)$" --- the space of internal degrees of a freedom of a string;
{\it the flag manifold of the Virasoro-Bott group\/} $\Vir$; the homogeneous
space
$\Diff_+(\circle)/\circle$ ($\Diff_+(\circle)$ is the group of diffeomorphisms
of a circle $\circle$ preserving an orientation); this space is identified via
Kirillov construction [13;A.A.Kirillov] with
\item"$S$" --- {\it the class of the univalent functions\/} $f(z)$ in
the unit complex disc $D_+$ ($D_+=\{z\in\Bbb C$: $|z|\le 1\}$) such that
$f(0)=0$, $f'(0)=1$, $f'(e^{it})\ne 0$; the natural coordinates on $S$ are
coefficients $c_k$ of the Taylor expansion of an univalent function $f(z)$:
$f(z)=z+c_1z^2+c_2z^3+\ldots+c_{n-1}z^n+c_nz^{n+1}+\ldots$.
\item"$\Cal C$" --- {\it the universal deformation of a complex disc\/}
with $M(\Vir)$ as a base and with fibers isomorphic to $D_+$.
The coordinates on $\Cal C$ are $z, c_1, c_2,\ldots, c_n,\ldots$, where $c_k$
are coordinates on the base and $z$ is a coordinate in the fibers.
\item"$M(\Vir)$"$\sq\,\,\, Q^*$ --- the space of both external and internal
degrees
of freedom of a string, the same as the bundle over $M(\Vir)$ associated with
$p:\Cal C\mapsto M(Vir)$, which fibers are $\Map(C/M(\Vir);\Bbb C^n)^*$ --
linear spaces dual to ones of mappings of fibers of
$p:\Cal C\mapsto M(\Vir)$ into $\Bbb C^n$.
\endroster

\subhead 2.1. Elements of Gervais-Matsuo differential $W$-geometry: classical
Toda fields in the complex analogue of Frenet theory; the mapping \linebreak
$\LDOP^{\Bbb C}(\circle)^{\reg}_+\mapsto\CVect(M(\Vir)\sq Q^*)$; Gervais-Matsuo
Lie
quasi(pseudo)al-\linebreak gebra $\CGM^{\Bbb C}_n$ of classical (infinitesimal)
W-transformations; Gervais-Mat-\linebreak suo Poisson algebra $\GM^{\Bbb C}_n$
and the
monomorphism $\GM^{\Bbb C}_n\mapsto\GD(\ssl(n,\Bbb C))$.
Infinite dimensional geometry of the flag manifold $M(\cLDOP(\circle)_+)\simeq
M(W^r_\infty)$ ($W_\infty=(W^r_\infty)^{\Bbb C}$) for the Lie algebra
$\cLDOP(\circle)_+$ or for the real form $W^r_\infty$ of the Pope-Shen-Romans
$W_\infty$-algebra \endsubhead

Let us consider a world-sheet of a string $\bold x:D_+\mapsto\Bbb C^n$ ($\bold
x=\{x^\mu=x^\mu(z)\}$), then using the complex version of Frenet theory [12]
one may introduce the associate mappings $\Gr_k\bold x:D_+\mapsto\Gr_k(\Bbb
C^n)$, where $\Gr_k(\Bbb C^n)$ is the Grassmannian of $k$-dimensional planes
in $\Bbb C^n$, or the associated mapping $\Fl\,\bold x:D_+\mapsto\Fl(\Bbb
C^n)$, where $\Fl(\Bbb C^n)$ is the space of full flags in $\Bbb C^n$.
If we consider the homogeneous coordinates or go to the projective space
$\CP^n$, then {\it the classical Toda fields\/} on $D_+$ will be identified
with K\"ahler potentials of the images of $\Gr_{k+1}\bold x(z)$ in
$\Gr_{k+1}(\Bbb C^{n+1})$.

It is convenient to introduce so-called {\it homogeneous KP-coordinates\/}
for the mapping $\bold x:D_+\mapsto\CP^n$.
Indeed, let us consider the functions $x^\mu([z])$, $[z]=[z^{(0)}, z^{(1)}=z,
z^{(2)},\ldots z^{(n)},\ldots]$ such that $D^l_z\bold
x([z])=\frac{\partial\bold
x([z])}{\partial z^{(l)}}$ and $\bold x([z])=\bold x(z)$ for $z^{(2)},
z^{(3)},\ldots\mathbreak z^{(n)},\ldots=0$, $z=z^{(1)}/z^{(0)}$.
So the homogeneous KP-coordinates $[z^{(0)}, z^{(1)},\ldots, z^{(n)}]$ may be
regarded as certain ones in the complex projective space $\CP^n$ defined in
the neibourhood of the world-sheet of a string.

Let now $\DOP^{\Bbb C}(\circle)^{\reg}$ be a subalgebra of $\DOP^{\Bbb
C}(\circle)$, which consists of operators possessing a regular continuation to
the unit complex disc $D_+$; $\DOP^{\Bbb C}(\circle)^{\reg}_+$ be the
intersection of $\DOP^{\Bbb C}(\circle)^{\reg}$ with $\DOP^{\Bbb
C}(\circle)_+$; $\LDOP^{\Bbb C}(\circle)^{\reg}$ and $\LDOP^{\Bbb
C}(\circle)^{\reg}_+$ be the corresponding commutator Lie algebras.

It should be mentioned that $\LDOP^{\Bbb C}(\circle)^{\reg}_+$ naturally acts
by differential operators on the space $Q^*$, considered as the space of
sections of a trivialized bundle over $D_+$ with fibers isomorphic to $\Bbb
C^n$).
One of the results of J.-L.Gervais and Y.Matsuo was that this action
is linearized (i.e. becomes an action by vector fields) in KP-coordinates.

\proclaim{Proposition 2} The action of the Lie algebra $\LDOP^{\Bbb
C}(\circle)^{\reg}_+$ by differential operators on $Q^*$ (as the space of
sections of a trivialized bundle over $D_+$ with fibers isomorphic to $\Bbb
C^n$) may be extended to
the action on $M(\Vir)\sq Q^*$ (as the space of sections of a trivialized
bundle over $\Cal C$ with fibers isomorphic to $\Bbb C^n$) by vector fields.
\endproclaim

\demo{Proof} It is necessary just to mention that KP-coordinate system is
a homogeneous version of the coordinates on the universal deformation
$\Cal C$ of a complex disc $D_+$, which are easily expressed via the standard
coordinates $z, c_1 ,c_2,\ldots c_n,\ldots$.
\enddemo

Let us now consider more systematically the structure of the action of the Lie
algebra $\LDOP^{\Bbb C}(\circle)^{\reg}_+$ on $Q^*$ by differential operators.

It is rather reasonable following J.-L.Gervais and Y.Matsuo to restrict
ourselves to a consideration of differential operators of order less or equal
to $n$.
It means that we shall deal with the quotient $\LDOP^{\Bbb
C}(\circle)^{\reg}_+/\LDOP^{\Bbb C}(\circle)^{\reg}_{\ge n+1}$ of the Lie
algebra $\LDOP^{\Bbb C}(\circle)^{\reg}_+$ of regular differential operators
without free terms by its subalgebra $\LDOP^{\Bbb C}(\circle)^{\reg}_{\ge
n+1}$ of regular differential operators without free terms, which are not
contained terms with $D^k$ ($1\le k\le n$), the commutator algebra of
associative algebra $\DOP^{\Bbb C}(\circle)^{\reg}_{\ge n+1}$ of such
operators.
Of course, such quotient is not a Lie algebra, because the subalgebra
$\LDOP^{\Bbb C}(\circle)^{\reg}_{\ge n+1}$ is not an ideal in $\LDOP^{\Bbb
C}(\circle)^{\reg}_+$.
Nevertheless, the strict sequence
$$\align 0\longrightarrow&\LDOP^{\Bbb C}(\circle)^{\reg}_{\ge
n+1}\longrightarrow\LDOP^{\Bbb C}(\circle)^{\reg}_+\longrightarrow\\
&\LDOP^{\Bbb C}(\circle)^{\reg}_+/\LDOP^{\Bbb C}(\circle)^{\reg}_{\ge n+1}
\longrightarrow 0\endalign$$
may be resolved.
The resolving mapping identifies the quotient\linebreak $\LDOP^{\Bbb
C}(\circle)^{\reg}_+/\LDOP^{\Bbb C}(\circle)^{\reg}_{\ge n+1}$ with
the subspace $\DOP^{\Bbb C}(\circle)^{\reg}_{+;\le n}$ of regular differential
operators of order less or equal to $n$ without free terms in
the Lie algebra $\LDOP^{\Bbb C}(\circle)^{\reg}_+$.

The object, which elements of $\DOP^{\Bbb C}(\circle)^{\reg}_{+;\le n}$ form
is, indeed, a Lie quasialgebra (in terminology of I.Batalin [32] or Lie
pseudoalgebra in terminology of M.V.Karasev and V.P.Maslov [33]).
Namely, the commutator of two elements of $\DOP^{\Bbb
C}(\circle)^{\reg}_{+;\le n}$ acting on $Q^*$ ({\it classical (infinitesimal)
$W$-transformations\/} [11]) may be expressed via other elements with
coefficients, which are extrinsic invariants of the world-sheet of a string
(curvature, torsions and their derivatives).
So the structure functions of classical (infinitesimal) $W$-transformations are
the functions on the space $Q^*$, the object, on which classical
(infinitesimal) $W$-transformations act.
Therefore, classical (infinitesimal) $W$-transformations form a Lie
quasi(pseudo)al-\linebreak gebra, which will be called {\it Gervais-Matsuo
quasi(pseudo)algebra\/} and will be denoted by $\CGM^{\Bbb C}_n$.

It should be marked (with respect to the possible constructing of classical
finite W-transformations) that Lie quasi(pseudo)algebras are infinitesimal
objects for Lie quasi(pseudo)groups of transformations in the finite
dimensional case [32,33].
Fixing a point of a manifold, on which the Lie quasi(pseudo)group of
transformations acts, one may receive a structure of a loop on the space of
the Lie quasi(pseudo)-\linebreak group.
In our case this loop will be just the loop of homogeneous space \linebreak
$\EXP(\LDOP^{\Bbb C}(\circle)^{\reg}_+)/\EXP(\LDOP^{\Bbb
C}(\circle)^{\reg}_{\ge n+1})$ defined by $\EXP(\DOP^{\Bbb
C}(\circle)^{\reg}_{+;\le n})$
as a resolution of the strict sequence
$$\align 0\longrightarrow&\EXP(\LDOP^{\Bbb C}(\circle)^{\reg}_{\ge
n+1})\longrightarrow\EXP(\LDOP^{\Bbb
C}(\circle)^{\reg}_+)\longrightarrow \\&\EXP(\LDOP^{\Bbb
C}(\circle)^{\reg}_+)/\EXP(\LDOP^{\Bbb C}(\circle)^{\reg}_{\ge
n+1})\longrightarrow 0\endalign$$
via {\it Sabinin construction\/} [34].
Nevertheless, infinite dimensional groups\linebreak $\EXP(\LDOP^{\Bbb
C}(\circle)^{\reg}_{\ge n+1})$ and $\EXP(\LDOP^{\Bbb C}(\circle)^{\reg}_+)$
(as I know) are not constructed and, therefore, the classical
finite $W$-transformations are not introduced; so Sabinin construction may be
considered only as an explanation of the appearance of Gervais-Matsuo
quasi(pseudo)algebra.
However, it should be marked, though the Lie quasi(pseudo)group of classical
finite $W$-transformations is not defined (though the author thinks that it is
possible) and, hence, the corresponding system of isotopic loops on it
can not be considered, the infinitesimal objects of such loops,
so called {\it Mikheev-Sabinin multialgebras\/} [35] may be derived only from
Gervais-Matsuo quasi(pseudo)algebra $\CGM^{\Bbb C}_n$.

It should be mentioned that $Q^*$ is a symplectic manifold and, therefore, one
may correspond to the generators of Gervais-Matsuo quasi(pseudo)algebra
$\CGM^{\Bbb C}_n$ their Hamiltonians, which are just functions of classical
Toda fields.
The structure of the Lie quasi(pseudo)algebra induces a structure of the
Poisson algebra on Hamiltonians, which will be called {\it
Gervais-Matsuo Poisson algebra\/} and will be denoted by $\GM^{\Bbb C}_n$.
There exists a very simple but remarkable fact, which we prefer to formulate
as a proposition.

\proclaim{Proposition 3} There exist a monomorphism of Poisson algebras
$$\GM^{\Bbb C}_n\mapsto\GD(\ssl(n,\Bbb C))$$
\endproclaim

Namely, Gervais-Matsuo algebra $\GM^{\Bbb C}_n$ is just a "regular" part of
Gelfand-Dickey algebra $\GD(\ssl(n,\Bbb C))$.

The significance of such monomorphism is explained by the fact that {\sl
Gervais-Matsuo differential $W$-geometry may be (at least, partially)
generalized on an arbitrary K\"ahler manifold as a complexified
target space}, because it is based essentially only on complex Frenet theory
(see [11,36]),
{\sl so one can construct analogs of Gervais-Matsuo quasi(pseudo)algebra
$\CGM^{\Bbb C}_n$ of classical (infinitesimal) $W$-transformations as well as
Gervais-Matsuo Poisson algebra $\GM^{\Bbb C}_n$ for an arbitrary K\"ahler
manifold.}

It is not less remarkable circumstance that {\sl Gervais-Matsuo differential
$W$-geo-\linebreak metry appeared being deeply connected with Sabinin program
of
"nonlinear geometric algebra" [37] and Weinstein-Karasev-Maslov approach to
nonlinear Poisson brackets [33]}; so from the point of view of algebraic
geometry Gervais-Matsuo $W$-geometry may be considered as
{\sl a penetration of nonassociative algebra into
the theory of embeddings of algebraic curves into K\"ahler varieties} (it is
reminiscent a little of a book of Yu.I.Manin [38]).

But let us now return to the main theme.
Above we obtained that the subalgebra $W^{\reg}_\infty$ of $W_\infty$-algebra,
generated by $V^s_m$, $m\ge 0$ acts by vector fields on the space $M(\Vir)\sq
Q^*$.
Nevertheless, we want to obtain an action of the whole $W_\infty$-algebra
(or $\cLDOP^{\Bbb C}(\circle)_+$) instead of $W^{\reg}_\infty$ (or
$\LDOP^{\Bbb C}(\circle)^{\reg}_+$) only.
So we should to enlarge the space of the internal symmetries of a string
by new degrees of freedom.
What we shall to do is to consider the flag manifold $M(\cLDOP(\circle)_+)$ for
the Lie algebra $\cLDOP(\circle)_+$ (or what is just the same
the flag manifold $M(W^r_\infty)$ of the real form $W^r_\infty$ of the
Pope-Shen-Romans $W_\infty$-algebra), to consider the squashed product
$M(W^r_\infty)\sq Q^*$ and to justify that $W_\infty$-algebra acts on
$M(W^r_\infty)\sq Q^*$ by vector fields.
To perform this program one needs in a detailed description of the flag
manifold $M(W^r_\infty)$.

The flag manifold $M(W^r_\infty)$ may be defined as a symplectic leaf of the
Poisson manifold $(W^r_\infty)^*$, the coadjoint module for the Lie algebra
$W^r_\infty$ (cf.[4]).
The tangent space of $M(W^r_\infty)$ at the initial point may be identified
with the quotient $W_\infty/W^{\reg}_\infty$, so the flag manifold
$M(W^r_\infty)$ is an almost complex manifold.

\proclaim{Proposition 4} The almost complex structure on the flag manifold
$M(W^r_\infty)$ is integrable.
\endproclaim

\demo{Proof} First of all, the almost complex structure on $M(W^r_\infty)$ is
formally integrable.
To prove that it is really integrable one should to construct an almost
complex embedding of $M(W^r_\infty)$ into some infinite dimensional complex
manifold.
The standard manifold for such purposes is an infinite dimensional
Grassmannian, f.e. one of subspaces in $W_\infty$, one of which is
$W^{\reg}_\infty$.
\enddemo

Being the symplectic leaf of the Poisson manifold $(W^r_\infty)^*$ the flag
space $M(W^r_\infty)$ possesses an infinite dimensional family of symplectic
structures $\omega_{\bold h,c}$, where $c$ corresponds to the central charge
of $W^r_\infty$ and $\bold h=(h_2, h_3,\ldots, h_n,\ldots)$ corresponds
to a character of subalgebra $W^{\reg}_\infty$.
Coupled with the complex structure 2-forms $\omega_{\bold h,c}$ define an
infinite dimensional family of (pseudo)K\"ahler metrics on the flag manifold
$M(W^r_\infty)$.

Each (pseudo)K\"ahler metric defines a prequantization bundle $E_{\bold
h,c}(M(W^r_\infty))$ over $M(W^r\infty)$, which is a Hermitean line bundle of
with the action of the Lie algebra $W_\infty$ by covariant derivatives with
curvature form $2\pi\omega_{\bold h,c}$.

In the Fock spaces $F(M(W^r_\infty),E_{\bold h,c})$ there are realized the
Verma modules over Pope-Shen-Romans $W_\infty$ algebra (Fock space of a pair
$(M,E)$ is a space dual to the space of sections of the bundle $E^*$ over $M$
[13: D.Juriev, Algebra Anal.]).
It should be mentioned that Verma modules over $\LDOP^{\Bbb C}(\circle)$
were investigated recently by V.Kac and A.Radul [9].\pagebreak

\subhead 2.2. $W$-symmetries of a second quantized free string (flat
background): $W$-ghosts, $W$-differential Banks-Peskin forms, Siegel
$W$-string fields and $W$-$\BRST$-operator in them; $W$-string Gauss-Manin
connection and $W$-string Kostant-Blattner-Sternberg pairings; geometrical
non-cancelation of Bowick-Rajeev $W$-anomaly --- absence of Bowick-Rajeev
$W$-vacua and gauge-invariant Siegel $W$-string fields; operator cancelation
of Bowick-Rajeev $W$-anomaly --- $W_N$-algebras \endsubhead

It should be mentioned that the flag manifold $M(W^r_\infty)$ for the
$W$-algebra $W^r_\infty$ admits an embedding into the infinite dimensional
analogue of the classical symmetric domain of type I [39].
Such embedding is defined by the mapping $W^r_\infty\mapsto\gl(\infty)$, where
$\gl(\infty)$ is the Lie algebra of linear operators in the space $\Cal
F(\circle)$ of functions on the circle $\circle$.
Because the infinite dimensional classical symmetric domain of type I admits a
representation as a space of complex structures on $\Cal
F(\circle)=\Map(\circle,\Bbb R)$ as well as on $\Map(\circle,\Bbb R^n)$, then
the obtain such representation for the flag manifold $M(W^r_\infty)$, too
(such representation is analogous to one of M.Bowick and S.Rajeev for the flag
manifold $M(\Vir)$ for the Virasoro-Bott group [40]).
Thus we have constructed the squash product $M(W^r_\infty)\sq Q^*$.

\proclaim{Proposition 5} The mapping $W^{\reg}_\infty\mapsto\CVect(M(\Vir)\sq
Q^*)$ is extended to the mapping $W_\infty\mapsto\CVect(M(W^r_\infty)\sq Q^*)$.
\endproclaim

This proposition is a specialization of a standard fact of the theory of
inductions to our infinite dimensional case.

Now we should to mention that Nambu-Goto action (the K\"ahler potential on
$Q^*$) defines the bundle $E_{\NG}(Q^*)$, which may be lifted to the bundle
$E_{\NG}(M(W^r_\infty)\sq Q^*)$; this procedure defines the first cohomology
class $H^1(W_\infty;\Cal O(M(W^r_\infty)\sq Q^*))$ of the Lie algebra
$W_\infty$ with coefficients in {\it "classical string fields"\/} $\Cal
O(M(W^r_\infty)\sq Q^*)$ (cf. [1], see also [30]).

The Hermitean line bundle $E_{\bold h,c}(M(W^r_\infty))$ over the flag
manifold $M(W^r_\infty)$ may be lifted to the bundle $E_{\bold
h,c}(M(W^r_\infty)\sq Q^*)$ over the space $M(W^r_\infty)\sq Q^*$, in which
the action of the Pope-Shen-Romans algebra $W_\infty$ with non-trivial central
charge is defined.
One may also include the first cohomology class of $W_\infty$ into such
action.
That means that one should consider the bundle $\tilde E_{\bold
h,c}(M(W^r_\infty)\sq Q^*)$, the tensor product of $E_{\bold
h,c}(M(W^r_\infty)\sq Q^*)$ and $E_{\NG}(M(W^r_\infty)\sq Q^*)$, instead of
$E_{\bold h,c}(M(W^r_\infty)\sq Q^*)$.

The Fock space $F(M(W^r_\infty)\sq Q^*),\tilde E_{\bold h,c}(M(W^r_\infty)\sq
Q^*))$ over
the pair $(M(W^r_\infty)\sq Q^*),\tilde E_{\bold h,c}(M(W^r_\infty)\sq Q^*))$
is just {\it "the configuration space for a second quantized free string
without ghosts after the accounting of $W$-symmetries"\/} (cf.[1], see also
[29]).

Our next purpose is to introduce $W$-ghosts, to consider
$W$-differential Banks-Peskin forms, Siegel $W$-string fields,
$W$-$\BRST$-operator in them and, thus, to construct {\it "the configuration
space for a second quantized free string with ghosts after the accounting
of $W$-symmetries"\/} in a way analogous to one of [1] (see also [29]).

Unfortunately, this program can not be performed even on the first step.
The deal is that one can not construct correctly the semi-infinite forms for
the Pope-Shen-Romans $W_\infty$-algebra (or for $\cLDOP^{\Bbb C}(\circle)_+$).
So we should to reduce our geometric picture to a finite order $n$.
Namely, in view of the (real) Radul mapping $\Cal
F((W^r_\infty)^*)\mapsto\GD(\ssl(n,\Bbb R))$ one may consider the flag manifold
$M(\GD(\ssl(n,\Bbb R)))$
for Gelfand-Dickey algebra instead of the flag manifold $M(W_\infty)$ for
$W_\infty$-algebra.
The flag manifold $M(\GD(\ssl(n,\Bbb R)))$ may be defined as a symplectic leaf
of
Gelfand-Dickey algebra $\GD(\ssl(n,\Bbb R))$.
The flag manifold $M(\GD(\ssl(n,\Bbb R))$ is a complex manifold as well as
$M(W^r_\infty)$
and possesses (at least, one [4]) $n$-parametric family of symplectic
structures $\omega_{\bold
h,c}$, where $\bold h=(h_2, \ldots h_n)$.
Coupled with the complex structure these 2-forms define $n$-parametric family
of (pseudo)K\"ahler metrics on the flag manifold $M(\GD(\ssl(n,\Bbb R)))$; each
(pseudo)K\"ahler metric defines a prequantization bundle $E_{\bold
h,c}(M(\GD(\ssl(n,\Bbb R))))$ over $M(\GD(\ssl(n,\Bbb R)))$.

\proclaim{Proposition 6} The mapping $W_\infty\mapsto\CVect(M(W^r_\infty)\sq
Q^*)$ may be reduced to the mapping $W_\infty\mapsto\CVect(M(\GD(\ssl(n,\Bbb
R)))\sq Q^*)$.
\endproclaim

\demo{Proof} It is just the consequence of the existence of Radul mapping.
\enddemo

One may perform for $M(\GD(\ssl(n,\Bbb R)))\sq Q^*$ all geometric procedures
described above
for $M(W^r_\infty)\sq Q^*$. Namely, the bundle $E_{\NG}(Q^*)$ may be lifted to
the bundle $E_{\NG}(M(\GD(\ssl(n,\Bbb R)))\sq Q^*)$; the Hermitean line bundle
$E_{\bold
h,c}(M(\GD(\ssl(n,\Bbb R))))$ may be lifted to the bundle $E_{\bold
h,c}(M(\GD(\ssl(n,\Bbb R)))\sq Q^*)$;
finally, one may consider the bundle $\tilde E_{\bold h,c}(M(\GD(\ssl(n,\Bbb
R)))\sq Q^*)$
as the tensor product of two previous ones.
It is natural consider the Fock space $F(M(\GD(\ssl(n,\Bbb R)))\sq Q^*,\tilde
E_{\bold
h,c}(M(\GD(\ssl(n,\Bbb R)))\sq Q^*))$ as {\it "the reduced configuration space
for a second
quantized free string without ghosts after the accounting of
$W$-symmetries"\/}.

But before a consideration of the corresponding ghosts it is reasonable to
consider the structure of action of the Lie algebra $W^{\Bbb C}_\infty$ on
$M(\GD(\ssl(n,\Bbb R)))\sq Q^*$ more systematically.

It seems that it is very convenient to restrict ourselves (as above) to a
consideration of differential operators of order less or equal to $n$.
It means that we shall deal with the quotient $\cLDOP^{\Bbb
C}(\circle)_+/\cLDOP^{\Bbb C}(\circle)_{\ge n+1}$ of the central extension of
the Lie algebra $\LDOP^{\Bbb C}(\circle)_+$ of differential operators without
free terms by the central extension of its subalgebra $\LDOP^{\Bbb
C}(\circle)_{\ge n+1}$ of differential operators without free terms, which are
not contained terms with $D^k$ ($1\le k\le n$). The corresponding strict
sequence
$$\align  0\longrightarrow\cLDOP^{\Bbb C}(\circle)_{\ge
n+1}\longrightarrow&\cLDOP^{\Bbb C}(\circle)_+\longrightarrow \\
&\cLDOP^{\Bbb C}(\circle)_+/\cLDOP^{\Bbb C}(\circle)_{\ge n+1}\longrightarrow
0\endalign$$
may be resolved. The resolving mapping identifies the quotient with the
subspace $\DOP^{\Bbb C}(\circle)_{+;\le n}$ of differential operators of order
less or equal to $n$ without free terms in the Lie algebra $\LDOP^{\Bbb
C}(\circle)_+$. The object, which elements od $\DOP^{\Bbb C}(\circle)_{+;\le
n}$ form is a Lie quasi(pseudo)algebra, which will be called {\it enlarged
Gervais-Matsuo quasi(pseudo)algebra\/} and will be denoted $\CGM^{\Bbb
C}_{n;(\enl)}$.
Gervais-Matsuo quasi(pseudo)algebra may be comprehended as "a less more than
one-half of" the enlarged Gervais-Matsuo quasi(pseudo)algebra.
It should be mentioned that the enlarged Gervais-Matsuo quasi(pseudo)algebra
is embedded in $\cLDOP^{\Bbb C}(\circle)_+$ rather than in $\LDOP^{\Bbb
C}(\circle)_+$ so we have obtained a central extension of the enlarged
Gervais-Matsuo quasi(pseudo)algebra, which will be denoted by
$\widehat{\CGM}^{\Bbb C}_{n;(\enl)}$.
If the enlarged Gervais-Matsuo quasi(pseudo)algebra $\CGM^{\Bbb C}_{n;(\enl)}$
acts on $M(\GD(\ssl(n,\Bbb R)))\sq Q^*$, then its central extension
$\widehat{\CGM}^{\Bbb C}_{n;(\enl)}$ acts in the line bundles over it.

So now we are able to introduce $W$-ghosts (related to the central extended
enlarged
Gervais-Matsuo quasi(pseudo)algebra), to consider the corresponding
$W$-differential Banks-Peskin forms and Siegel $W$-fields,
$W$-$\BRST$-operator in them and, thus, to construct {\it "the reduced
configuration space for a second quantized free string with ghosts after the
accounting of $W$-symmetries"\/}.
Namely, {\it $W$-ghosts\/} may be identified with elements of the
Lie quasi(pseudo)algebra $\CGM^{\Bbb C}_{n;(\enl)}$ acting on the manifold
$M(\GD(\ssl(n,\Bbb R)))\sq Q^*$ by vector fields, $W$-antighosts are dual
1-forms on
$M(\GD(\ssl(n,\Bbb R)))\sq Q^*$. {\it $W$-differential Banks-Peskin forms\/}
are just
differential forms on the manifold $M(\GD(\ssl(n,\Bbb R)))\sq Q^*$
valued in the line bundle
$\tilde E_{\bold h,c}$ generated by $W$-antighosts;
{\it semi-infinite $W$-differential Banks-Peskin forms\/} are
defined with respect to the $\Bbb Z$-grading on the Lie quasi(pseudo)algebra
$\CGM^{\Bbb C}_{n;(\enl)}$.

The space of $W$-differential Banks-Peskin forms should be denoted
by\linebreak
$\Omega_{\BP}(M(\GD(\ssl(n,\Bbb R)))\sq Q^*;\tilde E_{\bold h,c})$ and the
space of semi-infinite
$W$-differential \linebreak Banks-Peskin forms should be denoted by
$\Omega^{\SI}_{\BP}(M(\GD(\ssl(n,\Bbb R)))\sq Q^*;\tilde E_{\bold h,c})$.

\proclaim{Remark 1} The Lie quasi(pseudo)algebra $\widehat{\CGM}^{\Bbb
C}_{n;(\enl)}$ acts on the space \linebreak
$\Omega^{\SI}_{\BP}(M(\GD(\ssl(n,\Bbb R)))\sq Q^*;\tilde E_{\bold
h,c})$ with the central charge $c-2(2n^3-n-1)$.
\endproclaim

\definition{Definition 2}
\roster
\item"1." {\it Siegel $W$-string  fields\/} are elements of the space, dual to
the space\linebreak $\Omega^{\SI}_{\BP}(M(\GD(\ssl(n,\Bbb R)))\sq Q^*;\tilde
E_{\bold h,c})$ of
semi-infinite $W$-differential Banks-Pes-\linebreak kin forms.
\item"2." {\it $W$-$\BRST$-operator\/} is the operator $Q_{\BRST}$ in the
space of Siegel $W$-string fields dual to the exterior covariant derivative
$D$ in the space of semi-infinite $W$-differential Banks-Peskin forms.
\endroster
\enddefinition

\proclaim{Remark 2} $Q_{\BRST}^2=0$, iff $c=2(2n^3-n-1)$.
\endproclaim

Let us now study the aspects related to gauge-invariance of $W$-string fields
in sense of [1] (see also [31]).

Namely, the space $\Omega^{\SI}_{\BP}(M(\GD(\ssl(n,\Bbb R)))\sq Q^*;\tilde
E_{\bold h,c})$
of semi-infinite $W$-differential Banks-Peskin forms may be considered as a
space of holomorphic sections of a certain bundle over $M(\GD(\ssl(n,\Bbb
R)))$, which will
be called {\it Fock-plus-ghost bundle\/} and denoted by $\FG_{\bold
h,c}(M(\GD(\ssl(n,\Bbb R))))$.
Fibers of the vector bundle\linebreak $\FG_{\bold h,c}(M(\GD(\ssl(n,\Bbb R))))$
over points $x$ of the
flag manifold $M(\GD(\ssl(n,\Bbb R)))$ for Gelfand-Dickey algebra
$\GD(\ssl(n,\Bbb R))$ will be denoted
by\linebreak
$\left(\FG_{\bold h,c}\right)_x(M(\GD(\ssl(n,\Bbb R))))$.

There exists a set $\{P_x\}$ of natural gauge-fixing projectors $$P_x:\Cal
O(\FG_{\bold h,c}(M(\GD(\ssl(n,\Bbb R)))))\mapsto\left(\FG_{\bold
h,c}\right)_x(M(\GD(\ssl(n,\Bbb R))))$$
(here the flag manifold $M(\GD(\ssl(n,\Bbb R)))$ is interpreted as a space of
internal gauge
degrees of freedom, cf. [31]) and a set $\{I_x\}$ of embedding operators
$$I_x:\left(\FG_{\bold h,c}\right)_x(M(\GD(\ssl(n,\Bbb R))))\mapsto\Cal
O(\FG_{\bold
h,c}(M(\GD(\ssl(n,\Bbb R))))),$$
which obey two properties;\newline
$\qquad$ (1) $P_xI_x=\id$;\newline
$\qquad$ (2) $\left(\widehat{\CGM}^{\Bbb C}_{n;(\enl)}\right)_+(x)I_x=0$,
where $\left(\widehat{\CGM}^{\Bbb C}_{n;(\enl)}\right)_+(x)$ is the natural
resolution of the strict sequence
$$\align 0\longrightarrow&\left(\widehat{\CGM}^{\Bbb
C}_{n;(\enl)}\right)_0(x)\longrightarrow\widehat{\CGM}^{\Bbb
C}_{n;(\enl)}\longrightarrow \\
&\widehat{\CGM}^{\Bbb C}_{n;(\enl)}/\left(\widehat{\CGM}^{\Bbb
C}_{n;(\enl)}\right)_0(x)\longrightarrow 0,\endalign $$
where $\left(\widehat{\CGM}^{\Bbb
C}_{n;(\enl)}\right)_0(x)=\{v\in\widehat{\CGM}^{\Bbb C}_{n;(\enl)}: v(x)=0\}$.

One may define {\it $W$-string Gauss-Manin connection\/} in $\FG_{\bold
h,c}(M(\GD(\ssl(n,\Bbb R))))$ as
$$\align\nabla_v\varphi(x)=&\lim_{t\to\infty}t^{-1}(P_xI_{x+tv(x)}P_{x+tv(x)}\varphi-
P_x\varphi),\\  &v\in\Vect(M(\GD(\ssl(n,\Bbb R)))), x\in M(\GD(\ssl(n,\Bbb
R))).\endalign$$

This connection may be also define by the use of {\it $W$-string
Kostant-Blattner-Sternberg pairings\/}.
Namely, the Fock space $F(M(\GD(\ssl(n,\Bbb R))),\FG_{\bold h,c})$ possesses a
(pseudo)hermitean metric (cf.[13:D.Juriev, Algebra Anal.,31]).
If such metric is non-degenerate then it induces a metric $(\cdot,\cdot)$ in
the space $\Cal O(\FG_{\bold h,c}(M(\GD(\ssl(n,\Bbb R)))))$.
$W$-string Kostant-Blattner-Sternberg pairings are the mappings
$B_{x,y}(\cdot,\cdot)$ from the tensor product of
$\left(\FG_{\bold h,c}\right)_x(M(\GD(\ssl(n,\Bbb R))))$ and
$\left(\FG_{\bold h,c}\right)_y(M(\GD(\ssl(n,\Bbb R))))$ to $\Bbb C$, such that
$B_{x,y}(\varphi,\psi)=(I_x\varphi,I_y\psi)$.

The $W$-string Gauss-Manin connection $\nabla$ may be expressed via $W$-string
Kos-\linebreak tant-Blattner-Sternberg pairings in the following way:
$$\align\nabla_v\Phi(x)=0 &\text{ iff }
B_{x+tv(x),x}(\Phi(x+tv(x)),\psi)=B_{x,x}(\Phi(x),\psi)+o(t)\\ &\text{ for all
}
\psi\in\left(\FG_{\bold h,c}\right)_y(M(\GD(\ssl(n,\Bbb R)))),\endalign$$
where $\Phi(x)$ is a short notation for $P_x(\Phi)$.

\definition{Definition 3} (cf. [31]).
\roster
\item"1." A covariantly constant section of Fock-plus-ghost bundle\linebreak
$\FG_{\bold
h,c}(M(\GD(\ssl(n,\Bbb R))))$ over the flag manifold $M(\GD(\ssl(n,\Bbb R)))$
for
Gel-\linebreak fand-Dickey algebra
$\GD(\ssl(n,\Bbb R))$ is called {\it Bowick-Rajeev $W$-vacuum\/}.
\item"2." The space dual to the space of Bowick-Rajeev vacua is called {\it the
space of gauge-invariant Siegel $W$-string fields\/}.
\endroster
\enddefinition
\pagebreak

Unfortunately, Bowick-Rajeev $W$-vacua (or equivalently gauge-invariant Siegel
$W$-string fields) do not exist.
{\sl The phenomenon of geometric non-cancelation of Bowick-Rajeev $W$-anomaly
may be considered as an explanation of the fact that Gelfand-Dickey brackets
can not be quantized only by addition of central term.}
The global structural change of commutation relations is necessary.
Let us now describe the process of {\it operator cancelation\/} of
Bowick-Rajeev anomaly, which transforms Gelfand-Dickey algebras or enlarged
Gervais-Matsuo quasi(pseudo)algebras into $W_N$-algebras (it should be
mentioned that problem of quantization of Lie quasi(pseudo)algebras was
discussed in another context in the book of M.V.Karasev and V.P.Maslov [33]).
Of course, such operator cancelation does not supply us with $\BRST$-operator
for $W_N$-algebras.

Let us consider an arbitrary point $x$ of the flag manifold $M(\GD(\ssl(n,\Bbb
R)))$ for
Gelfand-Dickey algebra $\GD(\ssl(n,\Bbb R))$.
Let us embed the fiber\linebreak $\left(\FG_{\bold
h,c}\right)_x(M(\GD(\ssl(n,\Bbb R))))$ of
Fock-plus-ghost bundle $\FG_{\bold h,c}(M(\GD(\ssl(n,\Bbb R))))$\linebreak over
$x$ into the space
$\Cal O(\FG_{\bold h,c}(M(\GD(\ssl(n,\Bbb R)))))$ by use of $I_x$.
Now define the action of the real form $\widehat{\CGM}_{n;(\enl)}$ of enlarged
Gervais-Matsuo quasi(pseudo)algebra $\widehat{\CGM}_{n;(\enl)}^{\Bbb C}$ in
$V$ in the following way:
$$v(\phi)=P_x\nabla_vI_x\phi,\qquad \phi\in\left(\FG_{\bold
h,c}\right)_x(M(\GD(\ssl(n,\Bbb R)))), v\in\widehat{\CGM}_{n;(\enl)}.$$
It is not the action indeed because the commutation relations in the real form
of Gervais-Matsuo quasi(pseudo)algebra are broken, moreover, the object,
which we have received in no more a Lie quasi(pseudo)algebra, but an ordinary
algebra of operators.

It is just remarkable that the obtained algebras are just
(after a complexification) $W_N$-algebras of papers [3].

\head Conclusions \endhead

Thus, our plan is performed, the results may be summarized.

It was appeared that the main objects of the infinite dimensional $W$-geometry
of a second quantized free string are not infinite dimensional groups,
Lie algebras and their homogoneous spaces as it was in [1] but infinite
dimensional Lie quasi(pseudo)algebras (various modifications of Gervais-Matsuo
quasi(pseudo)-\linebreak algebra $\CGM^{\Bbb C}_n$ of classical (infinitesimal)
$W$-transformations), nonlinear Poisson \linebreak brackets and related
geometrical
structures.

As a consequence there exists a geometrical non-cancelation of Bowick-Rajeev
anomaly (the absence of gauge-invariant Siegel $W$-string fields).
Operator cancelation provides us with transformation of classical Lie
quasi(pseudo)algebras $\widehat{\CGM}^{\Bbb C}_{n;(\enl)}$ (central extended
enlarged Gervais-Matsuo algebras) into quantum $W_N$-algebras.

It should be mentioned that because realistic $W$-string field theory is
essentially the theory of self-interacting string field (see f.e. [41]), free
$W$-string
field theory may be considered only as a startpoint of the perturbation
approach to it.
Respectively, the infinite dimensional $W$-geometry of a second quantized free
string may be also regarded as a zero-approximation of the
noncommutative geometry of self-interacting $W$-string field.\pagebreak

\Refs
\roster
\item"[1]" D.Juriev, Infinite dimensional geometry and quantum field theory of
strings. I. Infinite dimensional geometry of a second quantized free string,
Alg. Groups Geom. 11 (1994).
\item"[2]" D.Juriev, Infinite dimensional geometry and quantum field theory of
strings. II. Infinite dimensional noncommutative geometry of self-interacting
string field, [in preparation].
\item"[3]" A.B.Zamolodchikov, Infinite additional symmetries in 2-dimensional
conformal field theory, Theor. Math. Phys. 65 (1985) 1205-1213;\newline
A.B.Zamolodchikov and V.A.Fateev, Conformal quantum field theory in two
dimensions having $\Bbb Z_3$-symmetry, Nucl. Phys. B 280 (1987)
644-660;\newline
S.L.Lukyanov and V.A.Fateev, Conformally invariant models in 2-dimensional
quantum field theory with $\Bbb Z_N$ symmetry, Zh. Exp. Theor. Fiz. 94(3)
(1988) 23-37;\newline
S.L.Lukyanov, Quantization of the Gel'fand-Dickey bracket, Funkt. Anal. Appl.
22 (1988) 255-263;\newline
J.-L.Gervais, Systematic approach to conformal theories, Nucl. Phys. B (Proc.
Suppl.) 5B (1988) 119-136;\newline
J.-L.Gervais and A.Bilal, Conformal theories with non-linearly-extended
Virasoro symmetries and Lie algebra classification. In: {\it Infinite
dimensional Lie algebras and Lie groups\/} Ed. V.Kac (World Scientific,
Singapore, 1989) 483-526;
Systematic approach to conformal field systems with
extended Virasoro symmetries, Phys. Lett. B 206 (1988) 412-420; Systematic
construction of conformal theories with higher-spin Virasoro symmetries, Nucl.
Phys. B 318 (1989) 579-630.
\item"[4]" V.Yu.Ovsienko and O.D.Ovsienko, Lie derivatives of order $n$ on the
line. Tensor meaning of the Gel'fand-Dickey bracket,  Adv. Soviet Math. 2
(1991)221-231;\newline
V.Yu.Ovsienko and B.A.Khesin, Symplectic leaves of the Gel'fand-Dickey
brackets and homotopy classes of nondegenerate curves, Funkt. Anal. Appl.
24 (1990) 33-40.
\item"[5]" O.S.Kravchenko and B.A.Khesin, A central extension of the algebra
of pseudodifferential symbols, Funkt. Anal. Appl. 25 (1991) 152-154.
\item"[6]" A.O.Radul, Lie algebras of differential operators, their central
extensions and $W$-algebras, Funkt. Anal. Appl. 25 (1991) 33-49;
Non-trivial central extensions of Lie algebras of differential operators in
two and higher dimensions, Phys. Lett. B 265 (1991) 86-91.
\item"[7]" I.Vaysburd and A.Radul, Differential operators and $W$-algebra,
Phys. Lett. B 274 (1992) 317-322.
\item"[8]" I.Bakas, B.Khesin and E.Kiritsis, The logarithm of the derivative
operator and higher spin algebras of $W_\infty$ type, Commun. Math. Phys. 151
(1993) 233-243.
\item"[9]" V.Kac and A.Radul, Quasiaffine highest weight modules over the Lie
algebra of differential operators on the circle, Commun. Math. Phys. 157
(1993) 429-458.
\item"[10]" B.Khesin and I.Zakharevich, Poisson-Lie group of
pseudodifferential operators and fractional KP-KdV hierarchies, C. R. Acad.
Sci. Paris I, 316 (1993) 621-626.
\item"[11]" J.-L.Gervais and Y.Matsuo, $W$-geometries, Phys. Lett. B 274
(1992) 309-316; Classical $A_n$-$W$-geometry, Commun. Math. Phys.
152 (1993) 317-368;\newline
J.-L.Gervais, $W$-geometry from chiral embeddings, J. Geom. Phys. 11 (1993)
293-304; Introduction to differential $W$-geometry, preprint LPTENS-93/38.
\item"[12]" P.Griffiths and S.Harris, {\it Principles of Algebraic Geometry\/}
(Wiley Interscience, New York, 1978).
\item"[13]" A.A.Kirillov, A K\"ahler structure on K-orbits of the group of
diffeomorphisms of a circle, Funkt. Anal. Appl. 21 (1987) 122-125;
\newline A.A.Kirillov and D.V.Juriev, The K\"ahler geometry on the infinite
dimensional homogeneous manifold $M=\Diff_+(\circle)/\Rot(\circle)$, Funkt.
Anal. Appl. 20 (1986) 322-324; The K\"ahler geometry on the infinite
dimensional homogeneous space $M=\Diff_+(\circle)/\Rot(\circle)$ Funkt. Anal.
Appl. 21 (1987) 284-293; Representations of the Virasoro algebra by
the orbit method, J. Geom. Phys. 5 (1988) 351-364 [reprinted in {\it Geometry
and Physics. Essays in honour of I.M.Gelfand.\/} Eds. S.G.Gindikin and
S.G.Singer (Pitagora Editrice, Bologna and Elsevier Sci. Publ., Amsterdam,
1991)];\newline
D.Juriev, The vocabulary of geometry and harmonic analysis on the
infinite-dimensional manifold $\Diff_+(\circle)/\circle$, Adv. Soviet Math. 2
(1991) 233-247; A model of Verma modules over the Virasoro algebra, Algebra
Anal. 2(2) (1990) 209-226; An infinite dimensional geometry of the universal
deformation of a complex disc. Russian J. Math. Phys. 2(1) (1994).
\item"[14]" D.Juriev, Quantum conformal field theory as infinite dimensional
noncommutative geometry, Russian Math.Surveys 46(4) (1991) 135-163.
\item"[15]" A.Yu.Morozov, Strings -- what are they? Uspekhi Fiz. Nauk 162(8)
(1992)
84-175; 162(11) (1992) 206.
\item"[16]" D.Juriev, Quantum projective field theory: quantum field analogs
of Euler formulas, Teor. Matem. Fiz. 92(1) (1992) 172-176; Quantum projective
field theory: quantum field analogs of Euler-Arnold equations in projective
$G$-hypermultiplets, Teor. Matem. Fiz. (1994).
\item"[17]" V.I.Arnold, {\it Mathematical methods of classical mechanics\/}
(Springer, 1976).
\item"[18]" E.Floratos and J.Iliopoulos, A note on the classical symmetries of
the closed bosonic membranes, Phys. Lett. B 201 (1988) 237-240;\newline
I.Antoniadis, P.Ditsas, E.Floratos and J.Iliopoulos, New realizations of the
Virasoro algebra as membrane symmetries, Nucl. Phys. B 300 (1988) 549-558.
\item"[19]" I.M.Gelfand and D.B.Fairlie, The algebra of Weyl symmetrized
polynomials and its quantum extension, Commun. Math. Phys. 136 (1991) 487-499.
\item"[20]" D.B.Fairlie, P.Fletcher and C.K.Zachos, Trigonometric structure
constants for new infi-\linebreak nite-dimensional algebras, Phys. Lett. B 218
(1989)
203-206;\newline
D.B.Fairlie and C.K.Zachos, Infinite dimensional algebras, sine brackets and
$\SU(\infty)$, Phys.Lett. B 224 (1989) 101-107;\newline
D.B.Fairlie, C.K.Zachos and P.Fletcher, Infinite dimensional algebras and a
trigonometric basis for the classical Lie algebras, J. Math. Phys. 31 (1990)
1088-1094.
\item"[21]" V.Kac and D.H.Peterson, Spin and wedge representations of
infinite-dimensional Lie algebras and groups, Proc. Nat'l Acad. Sci. USA 78
(1981) 3308-3312.
\item"[22]" D.B.Fairlie and J.Nuyts, Deformations and renormalisations of
$W_\infty$, Commun. Math. Phys. 134 (1990) 413-419.
\item"[23]" I.Bakas, The large-$N$ limit of extended conformal symmetries,
Phys. Lett. B 228 (1989) 57-63.
\item"[24]" C.N.Pope, X.Shen and L.J.Romans, The complete structure of
$W_\infty$, Phys. Lett. B 236 (1990) 173-178; A new higher-spin algebra and
the lone-star product, Phys. Lett. B 242 (1990) 401-406; $W_\infty$ and the
Racah-Wigner algebra, Nucl. Phys. B 339 (1990) 191-221.
\item"[25]"  I.Bakas, The structure of the $W_\infty$-algebra, Commun. Math.
Phys. 134 (1990) 487-508.
\item"[26]" I.N.Bernstein, I.M.Gelfand and S.I.Gelfand, Models of
representations of Lie groups. Trudy Semin. I.G.Petrovskogo 2 (1976) 3-21.
\item"[27]" B.L.Feigin, The Lie algebras $\gl(\lambda)$ and the cohomology of
the Lie algebra of differential operators, Russian Math. Surveys 35(2) (1988)
169-170.
\item"[28]" D.V.Juriev, Complex projective geometry and quantum projective
field theory, Teor. Matem. Fiz. (submitted).
\item"[29]" D.Juriev, Infinite dimensional geometry and closed string-field
theory, Lett. Math. Phys. 22 (1991) 1-6.
\item"[30]" D.Juriev, Cohomology of the Virasoro algebra with coefficients in
string fields. Lett. Math. Phys. 19 (1990) 355-356.
\item"[31]" D.Juriev, Gauge invariance of Banks-Peskin differential forms
(flat background), Lett. Math. Phys. 19 (1990) 59-64.
\item"[32]" I.Batalin, Quasigroup construction and first class constraints, J.
Math. Phys. 22 (1981) 1837-1850.
\item"[33]" M.V.Karasev and V.P.Maslov, {\it Nonlinear Poisson brackets.
Geometry and quantization\/} (Nauka, Moscow, 1991).
\item"[34]" L.V.Sabinin, On the equivalence of the category of loops and the
category of homogeneous spaces, Soviet Math. 13 (1972) 533-536.
\item"[35]" P.O.Mikheev and L.V.Sabinin, On the infinitesimal theory of local
analytic loops. Soviet Math. 36 (1988) 545-548.
\item"[36]" J.-L.Gervais and M.V.Saveliev, $W$-geometry of the Toda systems
associated with non-exceptional simple Lie algebras, preprint LPTENS-93/47.
\item"[37]" P.O.Mikheev and L.V.Sabinin, Smooth quasigroups and geometry.
VINITI Problems in Geometry 20 (1988) 75-110 [and several unpublished texts on
"nonlinear geometric algebra"].
\item"[38]" Yu.I.Manin, {\it Cubic forms\/} (Nauka, Moscow, 1969).
\item"[39]" E.Cartan, Sur les domaines born\'es homog\`enes de l'espace de $n$
variables complexes, Abh. Math. Seminar Hamburg 11 (1935) 116-162 [reprinted
in {\it Oeuvres Completes\/} (Gauthier-Villars,Paris, 1953)].
\item"[40]" M.J.Bowick and S.G.Rajeev, String theory as the K\"ahler geometry
of loop space, Phys. Lett. Rev. 58 (1987) 535-538; The holomorphic geometry of
closed bosonic string theory and $\Diff(\circle)/\circle$, Nucl. Phys. B 293
(1987) 348-384.
\item"[41]" A.Bilal and J.-L.Gervais, Non-linearly extended Virasoro algebras:
new prospects for building string theories, Nucl. Phys. B 326 (1989) 222-236.
\endroster
\endRefs
\enddocument